\documentclass[journal]{IEEEtran}
\ifCLASSINFOpdf
\else
\fi
\usepackage[utf8]{inputenc}
\usepackage{amsmath,amssymb,amsfonts}
\usepackage{cite}

\usepackage{algorithmic}
\usepackage{stfloats}
\usepackage{graphicx}
\usepackage{textcomp}
\usepackage{xcolor}
\usepackage{color}
\usepackage{pifont}
\usepackage{amsthm}

\usepackage{hyperref}
\usepackage{mathrsfs}
\usepackage{booktabs}
\usepackage{hyperref}
\usepackage{cases}
\usepackage{comment}
\usepackage{empheq} 
\usepackage{caption}
\usepackage{subcaption}
\allowdisplaybreaks[4]

\usepackage{url}  
\usepackage{graphicx}  
\usepackage[ruled,vlined,linesnumbered]{algorithm2e}

\usepackage{listings}

\begin{document}
\captionsetup[figure]{labelfont={rm},labelformat={default},labelsep=period,name={Fig.}}
\title{Delay Minimization for Movable Antennas-Enabled Anti-Jamming Communications With Mobile Edge Computing}
%
%
%

\author{Yue Xiu$^*$,~Yang Zhao$^*$,~Songjie Yang,~Minrui Xu,~Dusit Niyato~\IEEEmembership{Fellow,~IEEE},
~Yueyang Li,~Ning Wei~\IEEEmembership{Member,~IEEE}\\
\thanks{The corresponding author is Ning Wei.}
\thanks{$*$ These authors contributed equally to this work.}}

\maketitle
\begin{abstract}
In future 6G networks, anti-jamming will become a critical challenge, particularly with the development of intelligent jammers that can initiate malicious interference, posing a significant security threat to communication transmission. Additionally, 6G networks have introduced mobile edge computing (MEC) technology to reduce system delay for edge user equipment (UEs). Thus, one of the key challenges in wireless communications is minimizing the system delay while mitigating interference and improving the communication rate. However, the current fixed-position antenna (FPA) techniques have limited degrees of freedom (DoF) and high power consumption, making them inadequate for communication in highly interfering environments. To address these challenges, this paper proposes a novel MEC anti-jamming communication architecture supported by mobile antenna (MA) technology. The core of the MA technique lies in optimizing the position of the antennas to increase DoF. The increase in DoF enhances the system's anti-jamming capabilities and reduces system delay. In this study, our goal is to reduce system delay while ensuring communication security and computational requirements. We design the position of MAs for UEs and the base station (BS), optimize the transmit beamforming at the UEs and the receive beamforming at the BS, and adjust the offloading rates and resource allocation for computation tasks at the MEC server. Since the optimization problem is a non-convex multi-variable coupled problem, we propose an algorithm based on penalty dual decomposition (PDD) combined with successive convex approximation (SCA). The simulation results demonstrate that the proposed MA architecture and the corresponding schemes offer superior anti-jamming capabilities and reduce the system delay compared to FPA.
\end{abstract}


%
\IEEEpeerreviewmaketitle

\section{Introduction}
\IEEEPARstart{I}{n} future 6G communication systems, the inherent openness of wireless channels makes wireless transmissions susceptible to interference attacks\cite{6739367,5751298,7192640}. Therefore, interference mitigation has become an essential technique for 6G communication systems, aiming to prevent or mitigate intentional and unintentional interference to signal transmission and enhance the reliability of communication services. Furthermore, future 6G communication systems will increase mobile data, leading to increased storage and computational processing. However, introducing storage and computational processing functions will further increase system delay. Consequently, enhancing interference resistance while reducing system delay is a key challenge in future 6G communication systems\cite{10054381,8760275}.

Traditional interference mitigation techniques have focused on Frequency Hopping (FH)\cite{9733393,9108989}. FH is a well-established and extensively used method that allows wireless user equipment (UEs) to quickly change their operating frequency over various spectra, effectively mitigating potential interference threats\cite{8336971,8374072,7300436}. In \cite{8374072}, the authors designed a stochastic game to analyze the dynamics between jammers and legitimate UEs,  considering the jammer and transmitter as adversaries, each implementing optimal attack and defense strategies. In \cite{7300436}, the authors introduced an integrated approach combining rate adaptation and FH to mitigate interference from the jammer. This method enables the transmitter to bypass the jammer by altering its operating channel and adjusting its transmission rate. However, FH can lose effectiveness if a sophisticated jammer targets multiple channels simultaneously. In\cite{8336971}, the authors presented a pattern hopping technique that integrates traditional FH with pattern hopping to minimize error rates when faced with a jammer. Besides the FH, power control is another frequently employed technique. For example, in \cite{7744623}, the authors studied a wireless system impacted by a jammer, where the BS seeks to optimize transmission power to enhance system throughput while maintaining quality of service (QoS) standards for legitimate UEs. In\cite{8660628}, the authors developed an anti-jamming receiver to strengthen the communication system's anti-jamming. They also designed an optimal power control strategy to enhance achievable data rates.

To reduce system delay, recently emerging mobile edge computing (MEC) technology provides technical support to address the above problems. Specifically, MEC pushes computation to the network edge to support delay-sensitive applications on mobile devices\cite{9055737}. Numerous studies have explored using MEC technology to reduce system delay, with many considering the corresponding multi-user scenarios\cite{7762913,7956189,8387798,7307234,7542156,7572018}. In\cite{7762913}, the authors developed optimal resource allocation and offload strategies to minimize total energy consumption under computation delay constraints. In\cite{7956189}, the authors explored the allocation of transmission power and computing resources to UEs, using the penalty dual decomposition (PDD) method to optimize offloading decisions and resource distribution. In\cite{8387798}, the authors explored a multi-user time-division multiple access (TDMA) video compression offloading method, using cooperative communication and resource allocation to reduce system delays in local, edge, and partial offloading compression strategies. In\cite{7307234}, the authors introduced a decentralized game-theory-based algorithm for computation offloading in multiple-input and multiple-output (MIMO) scenarios. In \cite{7542156,7572018}, the authors focused on optimizing transmission power, CPU frequency, and task offloading rate in single-user MEC systems, to minimize UE delay while meeting energy budget constraints.

However, despite the effectiveness of interference mitigation schemes and MEC technology, the basis of these techniques is massive FPA. The massive deployment of FPA can lead to high hardware costs and system complexity. In addition, massive FPA is typically energy-intensive, requiring higher transmission and circuit power to enhance communication performance. A novel mobile antenna (MA) technique has recently been proposed to address these drawbacks\cite{10416363,10447471,feng2024movable,ding2024secure,tang2024secure}. This emerging technology shows promise in improving spectrum efficiency and reducing system power consumption and cost in 6G and other communication systems. Specifically, MA can adjust or move antenna positions according to requirements. This flexibility of the MA optimizes the position of the antenna for various communication environments, adapting to different signal propagation conditions.

MAs have been extensively used in wireless communication systems to improve security performance. In\cite{10416363,10447471,feng2024movable,ding2024secure,tang2024secure}, the authors explored enhancements in physical layer security for MA-assisted communication systems by jointly optimizing beamforming at the base station (BS) and the position of MAs to improve secrecy and transmission rates in the presence of eavesdroppers. Specifically, in\cite{10416363}, the authors explored physical layer security in MA-assisted systems, where information is sent from Alice, equipped with an MA, to a single antenna Bob while multiple single-antenna eavesdroppers are present. They optimized the beamforming at Alice and the position of the MA to maximize the achievable secrecy rate. In\cite{10447471}, a secure wireless system was proposed that supports MA. The beamforming at the transmitter and the antenna positions were jointly optimized under two criteria: power consumption minimization and secrecy rate maximization. In\cite{feng2024movable}, the authors examined a practical scenario in which Alice does not know the instantaneous non-line of sight component of the eavesdropping channel. They then minimized the probability of secrecy outage by jointly optimizing Alice's beamforming and antenna positions. In\cite{ding2024secure}, the physical layer security of MA-assisted full-duplex (FD) systems was studied. The authors maximized the total secrecy rate for uplink (UL) and downlink (DL) UEs by optimizing the BS's beamformer and MA positions. In\cite{tang2024secure}, the authors maximized the secrecy rate (SR) by simultaneously optimizing the transmit beamforming, artificial noise (AN) covariance matrix and MA positions, all within transmit power constraints and the minimum distance between MAs. In\cite{10304448,zhang2024movable,xiao2023multiuser,qin2023antenna}, the authors investigated MA's role in mitigating system interference. In\cite{10304448}, they modeled uplink multi-user channels to capture wireless channel variations induced by MA mobility at the BS. An optimization problem was then formulated to maximize the minimum achievable rate among multiple UEs in MA-assisted uplink communication by optimizing the receive beamforming and MA positions at the BS and UEs' transmit power. In\cite{zhang2024movable}, the authors proposed an MA-assisted multi-user hybrid beamforming scheme with a sub-connection structure, optimizing digital beamformers, analog beamformers, and sub-array positions under constraints of unit modulus, limited mobility region, and power budget to maximize the system's total rate. In\cite{xiao2023multiuser}, the study focused on the deployment of multiple MA in BS to improve multi-user communication performance. The researchers aimed to maximize the minimum achievable rate among UEs in MA-assisted uplink communication by jointly optimizing the positions of MAs, the BS's receive beamforming, and the transmit power of UEs to mitigate multi-user interference. In\cite{qin2023antenna}, the authors minimized system power by jointly optimizing the beamforming and position of MAs to reduce multiuser interference.

Let us briefly consider integrating MEC, anti-jamming, and MA technologies, which hold great potential:
\begin{itemize}
\item The application of MA technology will reduce the number of antennas in MEC systems and lower system costs.
\item MA technique introduces additional spatial DoFs, enhancing the efficiency of computational resource allocation in MEC systems with anti-jamming.
\item MA technique can extend the communication coverage of the MEC systems.
\end{itemize}
Although many authors have considered MEC and MA in existing references\cite{10416363,10447471,feng2024movable,ding2024secure,tang2024secure,10304448,zhang2024movable,xiao2023multiuser,qin2023antenna}, existing studies have not explored the use of MA to simultaneously reduce system delay in MEC offloading schemes and enhance interference mitigation strategies against jammers, where the jammer aims to degrade the quality of intended transmission by sending interference signals on legitimate channels. Therefore, this paper proposes an MA-assisted interference-resistant low-delay solution to improve the quality of wireless communication transmission and enhance the computational capability of communication networks. Specifically, the contributions of this paper are summarized as follows:
\begin{itemize}
\item In this paper, we investigate a novel MEC system model supported by MA with the presence of jammers. The system's UEs and BS are equipped with multiple antennas. Based on the MEC and anti-jamming communication framework, we derive the problem of jointly optimizing MA positions, beamforming, and resource allocation to minimize system delay under communication and computational constraints.
\item To address the nonlinear and non-convex delay minimization problem, we transform the original problem into an equivalent but more tractable form. Therefore, we propose a joint optimization algorithm based on successive convex approximation (SCA) and PDD frameworks to solve this problem. The proposed algorithm ensures convergence to a set of fixed solutions for the original optimization problem. Moreover, we show that the algorithm can be executed in parallel and provide a comprehensive analysis of its computational complexity.
\item To illustrate the advantages of MA technology in MEC systems with jammers, we present a comprehensive set of simulation results for various system configurations. The simulation results validate that the MA-assisted system significantly improves interference-resistant communication performance and reduces delay compared to traditional FPA systems. Furthermore, the results confirm the effectiveness of the proposed approach in suppressing interference and lowering system delay compared to existing methods.
\end{itemize}

\textbf{Organization:} The remainder of this paper is organized as follows: In \textbf{Section \ref{II}}, we introduce the system model for the multiple-antenna MA-enabled multi-user MEC system with a jammer and formulate the delay minimization problem. In \textbf{Section \ref{III}}, we develop a joint optimization algorithm for antenna position, beamforming design, and resource allocation, based on SCA and PDD frameworks. We also discuss the computational complexity of the proposed algorithm. In \textbf{Section \ref{VI}}, simulation results demonstrate the performance of the proposed algorithm in the new system. Finally, we conclude our findings in \textbf{Section \ref{V}}.

\section{System Model and Problem Formulation}\label{II}
\begin{figure}[htbp]
  \centering
  \includegraphics[width=0.43\textwidth, height=0.33\textwidth]{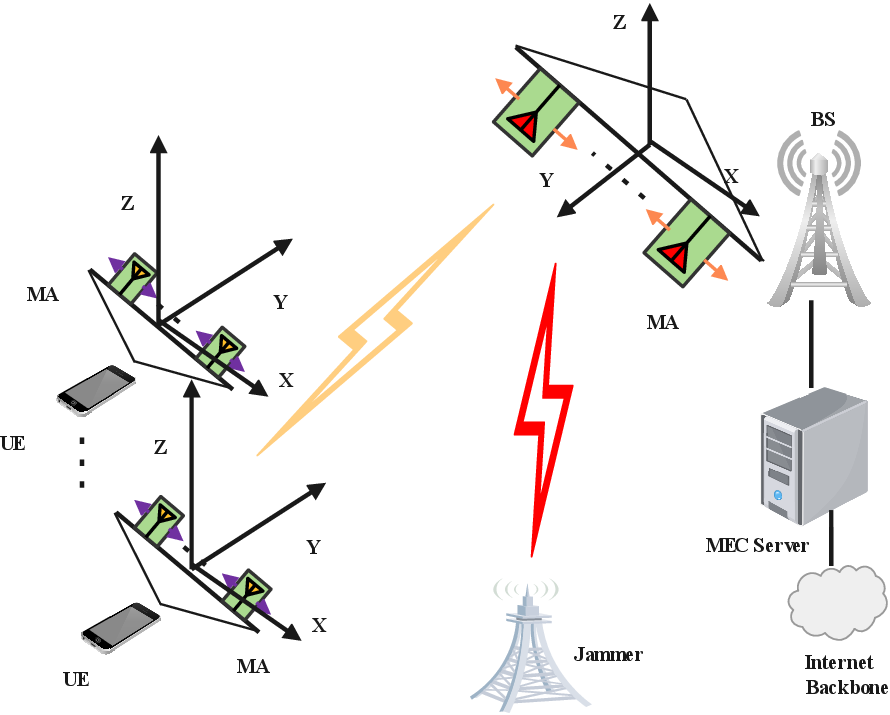}
  \captionsetup{justification=centering}
  \caption{ Illustration of a MEC communication system enabled by MA with a multi-antenna jammer.}
\label{FIGURE0}
\end{figure}
As shown in Fig.~\ref{FIGURE0}, the system comprises $K$ UEs, one BS, one MEC server, and one jammer. Each UE is equipped with $N_{t}$ MAs, and the BS is also equipped with $N_{r}$ MAs. Based on the Cartesian coordinate, the coordinates of the $n$-th antenna of the $k$-th UE are modeled as $\boldsymbol{p}_{k,n}=[x_{k,n},y_{k,n}]^{T}\in\mathbb{R}^{2\times1}\in\mathcal{C}_{t}$, $n\in\{1,\cdots,N_{t}\}$. The coordinates of the $m$-th antenna of the BS are represented as $\tilde{\boldsymbol{p}}_{b,r,m}=[\boldsymbol{p}_{r,m}^{T}, h_{b}]^{T}\in\mathbb{R}^{3\times 1}$, where $h_{b}$ denotes the height of the BS and $\boldsymbol{p}_{r,m}=[x_{r,m},y_{r,m}]^{T}\in\mathbb{R}^{2\times 1}\in\mathcal{C}_{b}$, $m\in\{1,\cdots,N_{r}\}$. Here, the mobile regions of the UEs' MAs are denoted as $\mathcal{C}_{t}$, and $\mathcal{C}_{r}$ represents the mobile region of the BS's MAs.

\subsection{Channel Model}
This paper assumes that the channel between the UE and the BS is a quasi-static block-fading model. For the uplink from the UE to the BS, the UE and the BS adjust the positions of MAs to update the channel. The coordinate set of the UE's MAs is denoted as $\boldsymbol{p}_{k}=[\boldsymbol{p}_{k,1},\boldsymbol{p}_{k,2},\cdots,\boldsymbol{p}_{k,N_{t}}]\in\mathbb{R}^{2\times N_{t}}$, and the BS's MAs is also denoted as $\boldsymbol{p}_{r}=[\boldsymbol{p}_{r,1},\boldsymbol{p}_{r,2},\cdots,\boldsymbol{p}_{r,N_{r}}]\in\mathbb{R}^{2\times N_{r}}$. The uplink channel from the $k$-th UE to the BS is represented as
\begin{align}
\boldsymbol{H}(\boldsymbol{p}_{k},\boldsymbol{p}_{r})=\boldsymbol{A}(\boldsymbol{p}_{r})^{H}\boldsymbol{\Sigma}_{k}\boldsymbol{A}(\boldsymbol{p}_{k})\in\mathbb{C}^{N_{r}\times N_{t}},\label{pro1}
\end{align}
where the array response matrix of transmit MAs $\boldsymbol{A}(\boldsymbol{p}_{k})$ is expressed as
\begin{align}
\boldsymbol{A}(\boldsymbol{p}_{k})=\left[\boldsymbol{a}(\boldsymbol{p}_{k,n})\right]\in\mathbb{C}^{L\times N_{t}}, \forall~n\in\{1,\cdots,N_{t}\},\label{pro2}
\end{align}
in which $\boldsymbol{a}(\boldsymbol{p}_{k,n})=\left[e^{j\frac{2\pi}{\lambda}\omega^{l}(\boldsymbol{p}_{k,n})}\right]^{T}\in\mathbb{C}^{L\times 1}$, $\forall~l\in\{1,\cdots,L\}$. $\lambda$ denotes the wavelength. Then, the difference in the propagation of the signal for the $l$-th transmit path between position $\boldsymbol{p}_{k,n}$ and the original coordinate is $\omega^{l}(\boldsymbol{p}_{k,n})=(\boldsymbol{\pi}_{k}^{l})^{T}\boldsymbol{p}_{k,n}=x_{k,n}\cos\vartheta_{k}^{l}\cos\psi_{k}^{l}+y_{k,n}\cos\vartheta_{k}^{l}\sin\psi_{k}^{l}$, where $\vartheta_{k}^{l}$ and $\psi_{k}^{l}$ are the azimuth AoD of the $l$-th transmit path given by $\psi_{k}^{l}\in[0,\pi]$, $\vartheta_{k}^{l}\in[0,\pi]$. Similarly, $\boldsymbol{A}(\boldsymbol{p}_{r})$ is expressed as
\begin{align}
\boldsymbol{A}(\boldsymbol{p}_{r})=[\boldsymbol{a}(\boldsymbol{p}_{r,m})]\in\mathbb{C}^{L\times N_{r}},\label{pro3}
\end{align}
where $\boldsymbol{a}(\boldsymbol{p}_{r,m})=\left[e^{j\frac{2\pi}{\lambda}\kappa^{l}(\boldsymbol{p}_{r,m})}\right]^{T}\in\mathbb{C}^{L\times 1}$, $\kappa^{l}(\boldsymbol{p}_{r,m})=(\boldsymbol{\pi}_{r}^{l})^{T}\boldsymbol{p}_{r,m}+h_{b}\sin\vartheta_{r}^{l}=x_{r,m}\cos\vartheta_{r}^{l}\cos\psi_{r}^{l}+y_{r,n}\cos\vartheta_{r}^{l}\sin\psi_{r}^{l}+h_{b}\sin\vartheta_{r}^{l}$, where $\psi_{r}^{l}$ and $\vartheta_{r}^{l}$ are the azimuth AoA of the $l$-th transmit path is given by $\psi_{r}^{l},\vartheta_{r}^{l}\in[0,\pi]$. According to \cite{10286328,wu2024movable,10458417}, since $\boldsymbol{H}(\boldsymbol{p}_{k},\boldsymbol{p}_{r})$ can be reconfigured and updated based on the MA positions, the channel is a function of both $\boldsymbol{p}_{k}$ and $\boldsymbol{p}_{r}$. 
Similarly, $\boldsymbol{H}_{J}$ is the channel from the smart jammer to the BS, and it is modeled as
\begin{align}
\boldsymbol{H}_{J}(\boldsymbol{p}_{r})=\boldsymbol{A}_{J}(\boldsymbol{p}_{r})^{H}\tilde{\boldsymbol{\Sigma}}\tilde{\boldsymbol{A}}_{J}\in\mathbb{C}^{N_{r}\times N_{J}},\label{pro4}
\end{align}
in which $\boldsymbol{A}_{J}(\boldsymbol{p}_{r})$ is the array response matrix, and it is a function of $\boldsymbol{p}_{r}$. Specifically, $\boldsymbol{A}_{J}(\boldsymbol{p}_{r})$ is expressed as
\begin{align}
\boldsymbol{A}_{J}(\boldsymbol{p}_{r})=[\boldsymbol{a}_{J}(\boldsymbol{p}_{r,m})]\in\mathbb{C}^{\tilde{L}\times N_{r}},
\end{align}
in which $\boldsymbol{a}_{J}(\boldsymbol{p}_{r,m})=\left[e^{j\frac{2\pi}{\lambda}\kappa_{J}^{l}(\boldsymbol{p}_{r,m})}\right]^{T}\in\mathbb{C}^{L\times 1}$, $\kappa_{J}^{l}(\boldsymbol{p}_{r,m})=(\boldsymbol{\pi}_{J}^{l})^{T}\boldsymbol{p}_{r,m}+h_{b}\sin\vartheta_{J}^{l}=x_{r,m}\cos\vartheta_{J}^{l}\cos\psi_{J}^{l}+y_{r,m}\cos\vartheta_{J}^{l}\sin\psi_{J}^{l}+h_{b}\sin\vartheta_{J}^{l}$, where $\psi_{J}^{l}$ and $\vartheta_{J}^{l}$ are the azimuth AoA of the $l$-th transmit path is given by $\psi_{J}^{l},\vartheta_{J}^{l}\in[0,\pi]$. 
Since fixed antennas are adopted in the jammer, the response matrix of the transmitted array $\tilde{\boldsymbol{A}}_{J}\in\mathbb{C}^{\tilde{L}\times N_{J}}$ is independent of the position variables $\boldsymbol{p}_{r}$, where $N_{J}$ is the number of transmitted antennas of the intelligent jammer.

\subsection{Signal Model}
Based on the channel model in (\ref{pro1}) and (\ref{pro4}), the received signal at the BS is expressed as
\begin{align}
\boldsymbol{y}_{r}=\sum\nolimits_{k=1}^{K}\boldsymbol{H}(\boldsymbol{p}_{k},\boldsymbol{p}_{r})\boldsymbol{x}_{k}+\boldsymbol{H}_{J}(\boldsymbol{p}_{r})\boldsymbol{z}+\boldsymbol{n}_{r},\label{pro6}
\end{align}
where $\boldsymbol{x}_{k}=\boldsymbol{w}_{k}s_{k}\in\mathbb{C}^{N_{t}\times 1}$ is the transmitted signal from the $k$-th UE. $\boldsymbol{z}\in\mathbb{C}^{N_{t}\times 1}$ denotes the jamming signal transmitted by the intelligent jammer to the BS, and the transmission power of $\boldsymbol{z}$ is represented by $P_{J}=\|\boldsymbol{z}\|^{2}$ and 
$\boldsymbol{n}_{r}\in\mathbb{C}^{N_{r}\times 1}\sim\mathcal{CN}(\boldsymbol{0},\sigma_{r}^{2}\boldsymbol{I})$. The received signal of the $k$-th UE after the processing at the BS is expressed as
\begin{align}
&\hat{s}_{k}=\boldsymbol{f}_{k}^{H}\boldsymbol{y}_{r}=\boldsymbol{f}^{H}_{k}\boldsymbol{H}(\boldsymbol{p}_{k},\boldsymbol{p}_{r})\boldsymbol{x}_{k}+\boldsymbol{f}^{H}_{k}\sum\nolimits_{k^{\prime}\neq k}^{K}\boldsymbol{H}(\boldsymbol{p}_{k^{\prime}},\boldsymbol{p}_{r})\nonumber\\
&\boldsymbol{x}_{k^{\prime}}+\boldsymbol{f}^{H}_{k}\boldsymbol{H}_{J}(\boldsymbol{p}_{r})\boldsymbol{z}+\boldsymbol{f}^{H}_{k}\boldsymbol{n}_{r},\label{pro7}
\end{align}
where $\boldsymbol{F}=[\boldsymbol{f}_{1},\cdots,\boldsymbol{f}_{K}]\in\mathbb{C}^{N_{r}\times K}$ denotes the received beamforming matrix at the BS. The communication achievable rate at the $k$-th UE is given in (\ref{pro8}) at the top of this page.
\begin{figure*}
\begin{align}
\Gamma_{k}=\log_{2}\left(1+\frac{|\boldsymbol{f}_{k}^{H}\boldsymbol{H}(\boldsymbol{p}_{k},\boldsymbol{p}_{r})\boldsymbol{w}_{k}|^{2}}{\sum\nolimits_{k^{\prime}\neq k}|\boldsymbol{f}_{k}^{H}\boldsymbol{H}(\boldsymbol{p}_{k^{\prime}},\boldsymbol{p}_{r})\boldsymbol{w}_{k^{\prime}}|^{2}+\|\boldsymbol{f}_{k}^{H}\|^{2}\sigma^{2}_{r}+\|\boldsymbol{f}_{k}^{H}\boldsymbol{H}_{J}(\boldsymbol{p}_{r})\boldsymbol{z}\|^{2}}\right).\label{pro8}
\end{align}
\hrulefill
\end{figure*}

\begin{table}[!ht]
\centering
\caption{Notations.}
\label{notations}
\begin{tabular}{cp{5.0cm}}
\toprule[1pt]
\textbf{Symbol} & \textbf{Descriptions} \\
\midrule
$K$ & The number of UEs. \\
$N_{t}/N_{r}/N_{J}$ &  The number of antennas. \\
$\boldsymbol{p}_{k,n}/\boldsymbol{p}_{r,m}$ & The coordinates of MAs. \\
$\mathcal{C}_{k,t}/\mathcal{C}_{u,r}/\mathcal{C}_{u,t}/\mathcal{C}_{b}$ & MA mobile region. \\
$\boldsymbol{H}(\boldsymbol{p}_{k},\boldsymbol{p}_{r})/\boldsymbol{H}_{J}(\boldsymbol{p}_{r})$ & The Channel between UE/jammer and BS. \\
$L/\tilde{L}$ & The number of channel paths. \\
$\boldsymbol{A}(\boldsymbol{p}_{k})/\tilde{\boldsymbol{A}}_{J}$ & Transmit MA field response matrix. \\
$\boldsymbol{A}(\boldsymbol{p}_{r})/\boldsymbol{A}_{J}(\boldsymbol{p}_{r})$ & Receive MA field response matrix. \\
$\vartheta_{J}^{\tilde{l}}/\vartheta_{k}^{l}/\vartheta_{r}^{l}$ & The AoAs. \\
$\psi_{J}^{\tilde{l}}/\psi_{k}^{l}/\psi_{r}^{l}$ & The AoDs. \\
$\boldsymbol{F}/\boldsymbol{w}_{k}$ & The transmit/receive beamforming. \\
$\boldsymbol{z}$ & The interference from the jammer. \\
$\boldsymbol{n}_{r}$ & The Noise sigal. \\
$\Delta_{k}$ & The number of computation task bits. \\
$\delta_{k}$ & The task offloading rate. \\
$\Psi_{1,k}/\Psi_{2,k}$ & computational resources/local computing. \\
$P_{J}/P_{k}$ & The jammer/UE transmit power. \\
$\Psi_{max}$ &  The computing budget of the MEC
server.\\
$\tilde{S}_{k}/\tilde{S}_{c,k}$ & The edge/local computational resource from UE. \\
$C_{k}$ & The system capacity. \\
\bottomrule[1pt]
\end{tabular}\label{TA1}
\end{table}

\subsection{Problem Formulation}
In the MA-aided anti-jamming communication system with MEC, we focus on the design of several key parameters to minimize the maximum delay among all UEs. The primary parameters include the beamforming vector 
$\boldsymbol{w}_{k}$ for the $k$-th UE, the beamforming vector $\boldsymbol{F}$ for the BS, the offloading rate $\delta_{k}$ for each UE, the computing resources $\delta_{k}$, and the locations of the MAs. By optimizing these parameters, we address the system's resource allocation issue. Specifically, the transmission beamforming vector $\boldsymbol{w}_{k}$ of the UE is designed to ensure efficient signal transmission from the UE to the BS, while the received beamforming vector $\boldsymbol{f}_{k}$ of the BS ensures accurate and timely processing of data received from the relay. The MA locations $\boldsymbol{p}_{r}$ for the BS and the $k$-th UE ensure that the relay is positioned optimally relative to the UE and BS, thereby maximizing coverage and reducing delay. The offloading rate 
$\delta_{k}$ balances the computational load between the UEs and the MEC server, optimizing processing time. Computing resources 
$\Psi_{1,k}$ are allocated based on the needs of each UE and the overall system capacity to achieve efficient task processing. Based on existing references\cite{7572018,7762913,8387798}, the mathematical formulation for minimizing system delay is as follows
\begin{subequations}
\begin{align}
\min_{\boldsymbol{p}_{k},\boldsymbol{p}_{r},\delta_{k},\atop{\atop{\Psi_{1,k},\boldsymbol{F},\boldsymbol{w}_{k}}}}&~\max_{k}\max\{\tilde{S}_{k},\tilde{S}_{c,k}\},\label{pro10a}\\
\mbox{s.t.}~
&\|\boldsymbol{w}_{k}\|^{2}\leq P_{k},&\label{pro10b}\\
&0\leq \delta_{k}\leq 1,&\label{pro10d}\\
&\sum\nolimits_{k=1}^{K}\Psi_{1,k}\leq \Psi_{max},&\label{pro10e}\\
&\boldsymbol{p}_{k}\in\mathcal{C}_{t},\boldsymbol{p}_{r}\in\mathcal{C}_{r},&\label{pro10f}\\
&\|\boldsymbol{p}_{k,n_{1}}-\boldsymbol{p}_{k,n_{2}}\|_{2}\geq d, n_{1}\neq n_{2},&\label{pro10g}\\
&\|\boldsymbol{p}_{r,m_{1}}-\boldsymbol{p}_{r,m_{2}}\|_{2}\geq d, m_{1}\neq m_{2},&\label{pro10i}
\end{align}\label{pro10}%
\end{subequations}
where the calculating the offloading time $\tilde{S}_{k}=\delta_{k}\Delta_{k}\left(\frac{1}{\Psi_{1,k}}+\frac{1}{B\Gamma_{k}}\right)$ and the local computation time
$\tilde{S}_{c,k}=\frac{(1-\delta_{k})\Delta_{k}}{\Psi_{2,k}}$
in objective function (\ref{pro10a}), in which we assume that the $k$-th UE is tasked with $\Delta_{k}$ bits, and this task can be partially offloaded, $\delta_{k}\in[0,1]$ represent the task offloading rate of the $k$-th UE, where $\delta_{k}\Delta_{k}$ bits are offloaded to the BS and processed by the MEC server, and the UE deals with a portion of $(1-\delta_{k})\Delta_{k}$ bits locally. The total edge computing offloading time comprises two elements: the time for transmitting data from the $k$-th UE to the relay and then to the BS and the processing time at the MEC server. $\Psi_{1,k}$ denotes the computational resources allocated to the MEC server for processing the $\sum\nolimits_{k=1}^{K}\Psi_{1,k}\leq \Psi_{max}$ bits, with $\Psi_{max}$ representing the computing budget of the MEC server. Let $\Psi_{2,k}$ be the local computing power of the $k$-th UE. The constraint (\ref{pro10b}) represents the transmission power limits of the UE. Constraint 
(\ref{pro10f}) defines the mobile areas for MA at the UE and BS. Constraints 
(\ref{pro10g}) and (\ref{pro10i}) are intended to prevent coupling effects between antennas, while $d$ denotes the minimum distance between each pair of antennas. The challenge of the problem (\ref{pro10}) arises from the highly non-convex nature of the channel vectors and achievable rates with respect to the MAs' position and beamforming. Furthermore, the coupling between matrix and vector variables complicates the problem (\ref{pro10}), making it difficult to obtain an optimal solution using existing optimization tools. Therefore, in the next section, we propose a suboptimal solution for the problem (\ref{pro10}). For simplicity, the symbols of this paper are summarized in \textbf{Table}.~\ref{TA1}.

\section{Proposed Joint Resource Allocation Algorithm}\label{III}
In this section, we first reformulate the problem (\ref{pro10}) into a more tractable equivalent form. We apply the PDD method to address highly coupled terms, introducing auxiliary variables and equality constraints that simplify the problem into a structure with multiple separable equality constraints. These constraints are then penalized and incorporated into the objective function, resulting in an augmented Lagrangian (AL) problem\cite{8691508}. The proposed PDD method consists of two iterative loops: the inner loop tackles the AL problem using an efficient algorithm based on the SCA method\cite{6159098, 8606437} within a BCD framework; the outer loop adjusts the dual variables or penalty parameters according to the level of constraint violation. Finally, we summarize the proposed algorithm and analyze its computational complexity.

To deal with problem (\ref{pro10}),
we ﬁrst introduce the following auxiliary variables $\gamma=\cdots=\gamma_{k}=\cdots=\gamma_{K}$ and
let $\gamma_{k}=\max\limits_{k}\max\{\tilde{S}_{k},\tilde{S}_{c,k}\}$, $\alpha_{1,k}=\max\limits_{k}\max\{\frac{\delta_{k}\Delta_{k}}{\Psi_{1,k}}\}$, $\alpha_{2,k}=\max\limits_{k}\max\{\frac{\delta_{k}\Delta_{k}}{B\Gamma_{k}}\}$, $\forall~k\in\mathcal{K}$. Thus, the objective function in (\ref{pro10}) can be transformed as $\min\limits_{\mathcal{X}}~\gamma$, where $\mathcal{X}=\{\boldsymbol{p}_{k},\boldsymbol{p}_{r},\Psi_{1,k},\boldsymbol{w}_{k},\delta_{k},\gamma,\gamma_{k},\Gamma_{k},$ $\alpha_{1,k},\alpha_{2,k}\}$. Since $\gamma_{k}=\max\limits_{k}\max\{\tilde{S}_{k},\tilde{S}_{c,k}\}$, ~$\max\limits_{k}\max\tilde{S}_{k}\leq \gamma_{k}$,~$\max\limits_{k}\max\tilde{S}_{c,k}\leq \gamma_{k}$. Then, since $\max\limits_{k}\max\tilde{S}_{k}$ can be rewritten as $\max\limits_{k}\max\tilde{S}_{k}=\max\limits_{k}\max\frac{\delta_{k}\Delta_{k}}{\Psi_{1,k}}+\max\limits_{k}\max\frac{\delta_{k}\Delta_{k}}{B\Gamma_{k}}$, $\max\limits_{k}\max\tilde{S}_{k}\leq\gamma_{k}$ is equivalently transformed into $\alpha_{1,k}+\alpha_{2,k}\leq\gamma_{k}$. Similarly, based on $\tilde{S}_{c,k}=\frac{(1-\delta_{k})\Delta_{k}}{\Psi_{2,k}}$, $\max\limits_{k}\max\tilde{S}_{c,k}\leq \gamma_{k}$ is updated $\tilde{S}_{c,k}=\frac{(1-\delta_{k})\Delta_{k}}{\Psi_{2,k}}\leq \gamma_{k}$.  The following new constraints are introduced
\begin{align}
&\alpha_{1,k}+\alpha_{2,k}\leq\gamma_{k}, \frac{\delta_{k}\Delta_{k}}{\Psi_{1,k}}\leq\alpha_{1,k},\frac{\delta_{k}\Delta_{k}}{B\Gamma_{k}}\leq\alpha_{2,k}, \nonumber\\
&\frac{(1-\delta_{k})\Delta_{k}}{\Psi_{2,k}}\leq \gamma_{k}, \gamma=\cdots=\gamma_{k}=\cdots=\gamma_{K}. \label{pro11}
\end{align}
Subsequently, the SINR expression in (\ref{pro8}) is highly complex and takes a fractional form, making the solution of the optimization problem extremely difficult. Specifically, the signal, interference, and noise powers in the SINR expression are determined by multiple coupled variables, further increasing the complexity of dealing with the SINR expression. To address this non-convex constraint, we introduce the auxiliary variable $\nu_{k}$. These variables help simplify the original SINR expression. Specifically, introducing an auxiliary variable $\nu_{k}$ allows the complex fractional form to be reformulated in a more tractable form, with $\Gamma_{k}$ being re-expressed as
\begin{align}
&\Gamma_{k}\leq\log_{2}(1+\nu_{k}),\label{pro12}
\end{align}
in which $\nu_{k}$ is given in (\ref{pro13}) at the top of this page.
\begin{figure*}
\begin{align}
\nu_{k}\leq \frac{|\boldsymbol{f}_{k}^{H}\boldsymbol{H}(\boldsymbol{p}_{k},\boldsymbol{p}_{r})\boldsymbol{w}_{k}|^{2}}{\sum\nolimits_{k^{\prime}\neq k}|\boldsymbol{f}_{k}^{H}\boldsymbol{H}(\boldsymbol{p}_{k^{\prime}},\boldsymbol{p}_{r})\boldsymbol{w}_{k^{\prime}}|^{2}+\|\boldsymbol{f}_{k}^{H}\|^{2}\sigma^{2}_{r}+\|\boldsymbol{f}_{k}^{H}\boldsymbol{H}_{J}(\boldsymbol{p}_{r})\boldsymbol{z}\|^{2}}.\label{pro13}
\end{align}
\hrulefill
\end{figure*}
To cope with the coupling constraints shown in (\ref{pro11}), (\ref{pro12}) and (\ref{pro13}), we
further introduce a set of auxiliary variables $\gamma_{k}$, $\tilde{\gamma}_{k}$, $\tilde{\alpha}_{1,k}$, $\tilde{\alpha}_{2,k}$, $\bar{\delta}_{k}$, $\hat{\delta}_{k}$, $\tilde{\delta}_{k}$, $\tilde{\Psi}_{1,k}$, $\tilde{\Gamma}_{k}$, $\nu_{k}$, $\tilde{\nu}_{k}$, $\tilde{u}_{k,k^{\prime}}=\boldsymbol{f}_{k}^{H}\tilde{\boldsymbol{\mu}}_{k^{\prime}}$, $\boldsymbol{\mu}_{k}=\boldsymbol{B}_{k}\tilde{\boldsymbol{w}}_{k}$, $\tilde{\boldsymbol{\mu}}_{k^{\prime}}=\boldsymbol{B}_{r}^{H}\boldsymbol{\Sigma}_{k}\boldsymbol{\mu}_{k^{\prime}}$, $\boldsymbol{w}_{k}=\tilde{\boldsymbol{w}}_{k}$, $\boldsymbol{f}_{k}=\tilde{\boldsymbol{f}}_{k}$, $\tilde{\boldsymbol{p}}_{k,n_{1}}$, $\tilde{\boldsymbol{p}}_{r,m_{1}}$, $\boldsymbol{u}_{k}=\boldsymbol{f}_{k}^{H}\boldsymbol{u}_{J}$, $\boldsymbol{u}_{J}=\boldsymbol{B}_{J}\tilde{\boldsymbol{\Sigma}}\tilde{\boldsymbol{A}}_{J}\boldsymbol{z}$, $\forall k,k^{\prime}\in \{1,\cdots,K\}$, 
where $\boldsymbol{B}_{k}=\boldsymbol{A}(\boldsymbol{p}_{k})$, $ \boldsymbol{B}_{r}=\boldsymbol{A}(\boldsymbol{p}_{r})$ and $ \boldsymbol{B}_{J}=\boldsymbol{A}_{J}(\boldsymbol{p}_{r})$ are auxiliary variables introduced to decouple the nonlinear relationship between the antenna positions and the array response matrix. Given that all elements in the matrix response must satisfy the constant modulus constraint, we introduce the constant modulus constraint for $|\boldsymbol{B}_{k}(l,n_{1})|=1$, $|\boldsymbol{B}_{r}(l,m_{1})|=1$, and $|\boldsymbol{B}_{J}(\tilde{l},m_{1})|=1$. Therefore, problem (\ref{pro10}) is rewritten as
\begin{subequations}
\begin{align}
\min_{\tilde{\mathcal{X}}}&~\gamma,\label{pro14a}\\
\mbox{s.t.}~
&\frac{\bar{\delta}_{k}\Delta_{k}}{\tilde{\Psi}_{1,k}}\leq\alpha_{1,k},&\label{pro14b}\\
&\frac{\tilde{\delta}_{k}\Delta_{k}}{B\Gamma_{k}}\leq \alpha_{2,k},&\label{pro14c}\\
&\tilde{\alpha}_{1,k}+\tilde{\alpha}_{2,k}\leq \gamma_{k}, \frac{(1-\hat{\delta}_{k})\Delta_{k}}{\Psi_{2,k}}\leq \tilde{\gamma}_{k},&\label{pro14d}\\
&\tilde{\Gamma}_{k}\leq\log_{2}(1+\nu_{k}), 0\leq\delta_{k}\leq 1,&\label{pro14e}\\
&\sum\nolimits_{k^{\prime}\neq k}|\tilde{u}_{k,k^{\prime}}|^{2}+\|\tilde{\boldsymbol{f}}_{k}\|^{2}\sigma_{r}^{2}+\|\boldsymbol{u}_{k}\|^{2}-\frac{|\tilde{u}_{k,k}|^{2}}{\tilde{\nu}_{k}}\leq 0, &\label{pro14f}\\
&\sum\nolimits_{k=1}^{K}\Psi_{1,k}\leq \Psi_{max},&\label{pro14g}\\
&\|\boldsymbol{w}_{k}\|^{2}\leq P_{k}, &\label{pro14h}\\
&\gamma=\gamma_{k}=\tilde{\gamma}_{k}, \alpha_{1,k}=\tilde{\alpha}_{1,k}, \alpha_{2,k}=\tilde{\alpha}_{2,k},&\nonumber\\ 
&\boldsymbol{w}_{k}=\tilde{\boldsymbol{w}}_{k}, \delta_{k}=\bar{\delta}_{k}=\hat{\delta}_{k}=\tilde{\delta}_{k}, \tilde{\Psi}_{1,k}=\Psi_{1,k}, \tilde{\Gamma}_{k}=\Gamma_{k},&\nonumber\\
&\tilde{\nu}_{k}=\nu_{k}, \tilde{\boldsymbol{f}}_{k}=\boldsymbol{f}_{k}, \tilde{u}_{k,k^{\prime}}=\boldsymbol{f}_{k}^{H}\tilde{\boldsymbol{\mu}}_{k^{\prime}},\tilde{\boldsymbol{\mu}}_{k^{\prime}}=\boldsymbol{B}_{r}^{H}\boldsymbol{\Sigma}_{k^{\prime}}\boldsymbol{\mu}_{k^{\prime}},&\nonumber\\
&\boldsymbol{\mu}_{k^{\prime}}=\boldsymbol{B}_{k^{\prime}}\tilde{\boldsymbol{w}}_{k^{\prime}},\boldsymbol{u}_{k}=\boldsymbol{f}_{k}^{H}\boldsymbol{u}_{J}, \boldsymbol{u}_{J}=\boldsymbol{B}_{J}\tilde{\boldsymbol{\Sigma}}\tilde{\boldsymbol{A}}_{J}\boldsymbol{z},&\label{pro14l}\\
&\tilde{\boldsymbol{p}}_{k,n_{1}}=\boldsymbol{p}_{k,n_{1}}-\boldsymbol{p}_{k,n_{2}},~\|\tilde{\boldsymbol{p}}_{k,n_{1}}\|_{2}\geq d,&\nonumber\\
&\tilde{\boldsymbol{p}}_{r,m_{1}}=\boldsymbol{p}_{r,m_{1}}-\boldsymbol{p}_{r,m_{2}},~\|\tilde{\boldsymbol{p}}_{r,m_{1}}\|_{2}\geq d,&\label{pro14_n}\\
&\boldsymbol{B}_{k}=\boldsymbol{A}(\boldsymbol{p}_{k}), \boldsymbol{B}_{r}=\boldsymbol{A}(\boldsymbol{p}_{r}), \boldsymbol{B}_{J}=\boldsymbol{A}_{J}(\boldsymbol{p}_{r}),&\label{pro14n}\\
&|\boldsymbol{B}_{k}(l,n_{1})|=1, |\boldsymbol{B}_{r}(l,m_{1})|=1, |\boldsymbol{B}_{J}(\tilde{l},m_{1})|=1,&\label{pro14o}
\end{align}\label{pro14}%
\end{subequations}
where $\tilde{\mathcal{X}}=\{$ $\gamma$, $\gamma_{k}$, $\tilde{\gamma}_{k}$, $\alpha_{1,k}$, $\alpha_{2,k}$, $\tilde{\alpha}_{1,k}$, $\tilde{\alpha}_{2,k}$, $\delta_{k}$, $\bar{\delta}_{k}$, $\hat{\delta}_{k}$, $\tilde{\delta}_{k}$, $\Psi_{1,k}$, $\tilde{\Psi}_{1,k}$, $\Gamma_{k}$, $\tilde{\Gamma}_{k}$, $\nu_{k}$, $\tilde{\nu}_{k}$, $u_{k,k^{\prime}}$, $\tilde{\boldsymbol{f}}_{k}$, $\boldsymbol{f}_{k}$, $\boldsymbol{\mu}_{k}$, $\tilde{\boldsymbol{\mu}}_{k^{\prime}}$, $\boldsymbol{u}_{k}$, $\boldsymbol{u}_{J}$, $\boldsymbol{w}_{k}$, $\tilde{\boldsymbol{w}}_{k}$, $\tilde{\boldsymbol{p}}_{k,n_{1}}$, $ \tilde{\boldsymbol{p}}_{r,m_{1}}$,  $\boldsymbol{B}_{k}$, $\boldsymbol{B}_{r}$, $\boldsymbol{B}_{J}$ $\}$.

\section{Proposed PDD-based Algorithm}
Based on the PDD framework, we move the equality constraints
(\ref{pro14l})-(\ref{pro14n}) into the objective function together with
Lagrange multipliers $\lambda_{\gamma_{k}}$, $\lambda_{\tilde{\gamma}_{k}}$, $\lambda_{\alpha_{1,k}}$, $\lambda_{\alpha_{2,k}}$, $\lambda_{\bar{\delta}_{k}}$, $\lambda_{\hat{\delta}_{k}}$, $\lambda_{\tilde{\delta}_{k}}$, $\lambda_{\Psi_{1,k}}$, $\lambda_{\Gamma_{k}}$, $\lambda_{\tilde{\nu}_{k}}$, $\lambda_{u_{k,k^{\prime}}}$,  $\boldsymbol{\lambda}_{\boldsymbol{f}_{k}}$, $\boldsymbol{\lambda}_{\boldsymbol{\mu}_{k}}$, $\boldsymbol{\lambda}_{\tilde{\boldsymbol{\mu}}_{k^{\prime}}}$, $\boldsymbol{\lambda}_{\boldsymbol{u}_{k}}$, $\boldsymbol{\lambda}_{\boldsymbol{u}_{J}}$, $\boldsymbol{\lambda}_{\boldsymbol{w}_{k}}$, $\boldsymbol{\lambda}_{\tilde{\boldsymbol{p}}_{k,n_{1}}}$, $ \boldsymbol{\lambda}_{\tilde{\boldsymbol{p}}_{r,m_{1}}}$,  $\boldsymbol{\lambda}_{\boldsymbol{B}_{k}}$, $\boldsymbol{\lambda}_{\boldsymbol{B}_{r}}$, and $\boldsymbol{\lambda}_{\boldsymbol{B}_{J}}$. The resulting AL problem can be given in (\ref{pro15}) at the top of the next page.
\begin{figure*}
\begin{subequations}
\begin{align}
\min_{\tilde{\mathcal{U}}}&~\gamma+\frac{1}{2\kappa}(\sum\nolimits_{k=1}^{K}(|\gamma-\gamma_{k}+\kappa\lambda_{\gamma_{k}}|^{2}+|\gamma-\tilde{\gamma}_{k}+\kappa\lambda_{\tilde{\gamma}_{k}}|^{2}+|\alpha_{1,k}-\tilde{\alpha}_{1,k}+\kappa\lambda_{\alpha_{1,k}}|^{2}+|\alpha_{2,k}-\tilde{\alpha}_{2,k}+\kappa\lambda_{\alpha_{2,k}}|^{2}&\nonumber\\
&+|\delta_{k}-\bar{\delta}_{k}+\kappa\lambda_{\bar{\delta}_{k}}|^{2}+|\delta_{k}-\hat{\delta}_{k}+\kappa\lambda_{\hat{\delta}_{k}}|^{2}+|\delta_{k}-\tilde{\delta}_{k}+\kappa\lambda_{\tilde{\delta}_{k}}|^{2}+|\Psi_{1,k}-\tilde{\Psi}_{1,k}+\kappa\lambda_{\Psi_{1,k}}|^{2}+|\Gamma_{k}-\tilde{\Gamma}_{k}+\kappa\lambda_{\Gamma_{k}}|^{2}&\nonumber\\
&+|\nu_{k}-\tilde{\nu}_{k}+\kappa\lambda_{\tilde{\nu}_{k}}|^{2}+\sum\nolimits_{k^{\prime}=1}^{K}(|\tilde{u}_{k,k^{\prime}}-\boldsymbol{u}_{k}\tilde{\boldsymbol{\mu}}_{k^{\prime}}+\kappa\lambda_{\tilde{u}_{k,k^{\prime}}}|^{2}+\|\boldsymbol{\mu}_{k^{\prime}}-\boldsymbol{B}_{k^{\prime}}\tilde{\boldsymbol{w}}_{k^{\prime}}+\kappa\boldsymbol{\lambda}_{\boldsymbol{\mu}_{k^{\prime}}}\|^{2}+\|\tilde{\boldsymbol{\mu}}_{k^{\prime}}-\boldsymbol{B}_{r}^{H}\boldsymbol{\Sigma}_{k}\boldsymbol{\mu}_{k^{\prime}}\nonumber&\nonumber\\
&+\kappa\boldsymbol{\lambda}_{\tilde{\boldsymbol{\mu}}_{k^{\prime}}}\|^{2})
+\|\boldsymbol{w}_{k}-\tilde{\boldsymbol{w}}_{k}+\kappa\boldsymbol{\lambda}_{\boldsymbol{w}_{k}}\|^{2}+\|\boldsymbol{f}_{k}-\tilde{\boldsymbol{f}}_{k}+\kappa\boldsymbol{\lambda}_{\boldsymbol{f}_{k}}\|^{2}+\|\boldsymbol{u}_{k}-\boldsymbol{f}_{k}^{H}\boldsymbol{u}_{J}+\kappa\boldsymbol{\lambda}_{\boldsymbol{u}_{k}}\|^{2} +\|\boldsymbol{u}_{J}-\boldsymbol{B}_{J}\tilde{\boldsymbol{\Sigma}}\tilde{\boldsymbol{A}}_{J}\boldsymbol{z}+\kappa\boldsymbol{\lambda}_{\boldsymbol{u}_{J}}\|^{2}
&\nonumber\\
&+\sum\nolimits_{n_{1}\neq n_{2}}^{N_{u}}|\tilde{\boldsymbol{p}}_{k,n_{1}}-(\boldsymbol{p}_{k,n_{1}}-\boldsymbol{p}_{k,n_{2}})+\kappa\boldsymbol{\lambda}_{\tilde{\boldsymbol{p}}_{k,n_{1}}}|^{2}+\sum\nolimits_{m_{1}\neq m_{2}}^{N_{r}}|\tilde{\boldsymbol{p}}_{r,m_{1}}-(\boldsymbol{p}_{r,m_{1}}-\boldsymbol{p}_{r,m_{2}})+\kappa\boldsymbol{\lambda}_{\tilde{\boldsymbol{p}}_{r,m_{1}}}|^{2}+&\nonumber\\
&\|\boldsymbol{A}(\boldsymbol{p}_{k})-\boldsymbol{B}_{k}+\kappa\boldsymbol{\lambda}_{\boldsymbol{B}_{k}}\|^{2})+\|\boldsymbol{A}(\boldsymbol{p}_{r})-\boldsymbol{B}_{r}+\kappa\boldsymbol{\lambda}_{\boldsymbol{B}_{r}}\|^{2}+\|\boldsymbol{A}_{J}(\boldsymbol{p}_{r})-\boldsymbol{B}_{J}+\kappa\boldsymbol{\lambda}_{\boldsymbol{B}_{J}}\|^{2},&\label{pro15a}\\
\mbox{s.t.}~
&(\ref{pro14b})-(\ref{pro14h}),(\ref{pro14o}),&\label{pro15b}\\
&\|\tilde{\boldsymbol{p}}_{k,n_{1}}\|_{2}\geq d,~\|\tilde{\boldsymbol{p}}_{r,m_{1}}\|_{2}\geq d.&\label{pro15c}
\end{align}\label{pro15}
\end{subequations} 
\hrulefill
\end{figure*}

\subsection{Proposed SCA-based Algorithm for dealing with non-convex in problem (\ref{pro15})}
Due to the remaining non-convex constraints, problem (\ref{pro15}) remains a non-convex problem. Next, we address these non-convex constraints (\ref{pro14b}), (\ref{pro14c}) and (\ref{pro14d}). According to\cite{9348933,10035459}, the SCA algorithm can tackle these non-convex constraints. Specifically, by linearizing them using a first-order Taylor series approximation, we approximate these constraints as convex constraints. Subsequently, we solve the AL problem using an iterative algorithm based on SCA. The core idea of the SCA algorithm is to transform the non-convex constraints into a series of convex subproblems that are easier to solve through local convexification. This process typically involves expanding the non-convex functions using a first-order Taylor series, approximating them as linear functions in the neighborhood of the current solution, thereby enabling the application of convex optimization techniques.

First, let us focus on constraints (\ref{pro14b}), (\ref{pro14c}), and (\ref{pro14d}). Constraints (\ref{pro14b}), (\ref{pro14c}), and (\ref{pro14d}) can be rewritten in the following form
\begin{align}
&\bar{\delta}_{k}^{2}+\frac{\Delta_{k}^{2}}{\tilde{\Psi}_{1,k}^{2}}\leq 2\alpha_{1,k},
\tilde{\delta}_{k}^{2}+\frac{\Delta_{k}^{2}}{B^{2}\Gamma_{k}^{2}}\leq 2\alpha_{2,k}, (1-\hat{\delta}_{k})^{2}+\frac{\Delta_{k}^{2}}{\Psi_{2,k}^{2}}\leq\nonumber\\
&\tilde{\gamma}_{k}, \tilde{\Gamma}_{k}-\log_{2}(1+\nu_{k}^{(t-1)})+\frac{1}{2\ln2(1+\nu_{k}^{(t-1)})}(\nu_{k}-\nu_{k}^{(t-1)}).
\label{pro16}
\end{align}
Then, deal with constraint (\ref{pro14f}). Constraint (\ref{pro14f}) can be rewritten in the following form
\begin{align}
&\sum_{k^{\prime}\neq k}|\tilde{u}_{k,k^{\prime}}|^{2}+\|\tilde{\boldsymbol{f}}_{k}\|^{2}\sigma^{2}_{r}+\|\boldsymbol{u}_{k}\|^{2}-(\tilde{u}_{k,k}^{(t-1),*}\tilde{u}_{k,k}/\tilde{\nu}_{k}^{(t-1)}+\nonumber\\
&\tilde{u}_{k,k}^{*}\tilde{u}_{k,k}^{(t-1)}/\tilde{\nu}_{k}^{(t-1)}-|\tilde{u}_{k,k}^{(t-1)}|^{2}\tilde{\nu}_{k}/(\tilde{\nu}_{k}^{(t-1)})^{2})\leq 0.\label{pro17}
\end{align}
Finally, we cope with constraint (\ref{pro15c}) and $\boldsymbol{t}_{k,m_{2}}$,$\boldsymbol{u}_{r,n_{2}}$ are given as initial MA positions. Constraint (\ref{pro15c}) can be rewritten in the following form
\begin{align}
&\|\tilde{\boldsymbol{p}}_{k,n_{1}}^{(t-1)}\|_{2}-2\tilde{\boldsymbol{p}}_{k,n_{1}}^{(t-1)}\tilde{\boldsymbol{p}}_{k,n_{1}}^{T}\geq d,\nonumber\\
&\|\tilde{\boldsymbol{p}}_{r,m_{1}}^{(t-1)}\|_{2}-2\tilde{\boldsymbol{p}}_{r,m_{1}}^{(t-1)}\tilde{\boldsymbol{p}}_{r,m_{1}}^{T}\geq d,\label{pro18}
\end{align}
where $\tilde{\boldsymbol{p}}_{k,n_{1}}^{(t-1)}=\boldsymbol{p}_{k,n_{2}}$, $\tilde{\boldsymbol{p}}_{r,m_{1}}^{(t-1)}=\boldsymbol{p}_{r,m_{2}}$.
Let us deﬁne the constraint violation $\|\boldsymbol{h}(\mathcal{X}^{(t)})\|_{\infty}$ as (\ref{pro_18}) on the next page.
\begin{figure*}
\begin{align}
&\|\boldsymbol{h}(\mathcal{X}^{(t)})\|_{\infty}=\max\{|\gamma-\gamma_{k}|,|\gamma-\tilde{\gamma}_{k}|,|\alpha_{1,k}-\tilde{\alpha}_{1,k}|,|\alpha_{2,k}-\tilde{\alpha}_{2,k}|,|\delta_{k}-\bar{\delta}_{k}|,|\delta_{k}-\hat{\delta}_{k}|,|\delta_{k}-\tilde{\delta}_{k}|,|\Psi_{1,k}-\tilde{\Psi}_{1,k}|,\nonumber\\
&|\Gamma_{k}-\tilde{\Gamma}_{k}|,|\nu_{k}-\tilde{\nu}_{k}|,|\tilde{u}_{k,k^{\prime}}-\boldsymbol{u}_{k}\tilde{\boldsymbol{\mu}}_{k^{\prime}}|,\|\boldsymbol{\mu}_{k^{\prime}}-\boldsymbol{B}_{k^{\prime}}\tilde{\boldsymbol{w}}_{k^{\prime}}\|,
\|\tilde{\boldsymbol{\mu}}_{k^{\prime}}-\boldsymbol{B}_{r}^{H}\boldsymbol{\Sigma}_{k}\boldsymbol{\mu}_{k^{\prime}}\|, \|\boldsymbol{w}_{k}-\tilde{\boldsymbol{w}}_{k}\|, \|\boldsymbol{f}_{k}-\tilde{\boldsymbol{f}}_{k}\|, \|\boldsymbol{u}_{k}-\boldsymbol{f}_{k}^{H}\boldsymbol{u}_{J}\|,\nonumber\\
&\|\boldsymbol{u}_{J}-\boldsymbol{B}_{J}^{H}\tilde{\boldsymbol{\Sigma}}\tilde{\boldsymbol{A}}_{J}\boldsymbol{z}\|, \|\tilde{\boldsymbol{p}}_{k,n_{1}}-(\boldsymbol{p}_{k,n_{1}}-\boldsymbol{p}_{k,n_{2}})\|, \|\tilde{\boldsymbol{p}}_{r,m_{1}}-(\boldsymbol{p}_{r,m_{1}}-\boldsymbol{p}_{r,m_{2}})\|, \|\boldsymbol{A}(\boldsymbol{p}_{k})-\boldsymbol{B}_{k}\|,\|\boldsymbol{A}(\boldsymbol{p}_{r})-\boldsymbol{B}_{r}\|,\nonumber\\
&\|\boldsymbol{A}_{J}(\boldsymbol{p}_{r})-\boldsymbol{B}_{J}\|,~\forall~k,k^{\prime}.\label{pro_18}
\end{align}
\hrulefill
\end{figure*}

Following the execution framework of the PDD-based algorithm, the penalty parameter $\kappa$ is updated using the formula $\kappa^{(t+1)}=c\kappa^{(t)}$, $(0 < c < 1)$, depending on the level of constraint violation, and the dual variables are adjusted based on (\ref{pro19}) at the top of the next page,
\begin{figure*}
\begin{align}
&\lambda_{\gamma_{k}}^{(t+1)}=\lambda_{\gamma_{k}}^{(t)}+\frac{1}{\kappa^{(t)}}(\gamma-\gamma_{k}),\lambda_{\tilde{\gamma}_{k}}^{(t+1)}=\lambda_{\tilde{\gamma}_{k}}^{(t)}+\frac{1}{\kappa^{(t)}}(\gamma-\tilde{\gamma}_{k}),\lambda_{\alpha_{1,k}}^{(t+1)}=\lambda_{\alpha_{1,k}}^{(t)}+\frac{1}{\kappa^{(t)}}(\alpha_{1,k}-\tilde{\alpha}_{1,k}),\lambda_{\alpha_{2,k}}^{(t+1)}=\lambda_{\alpha_{2,k}}^{(t)}+\frac{1}{\kappa^{(t)}}(\alpha_{2,k}-\tilde{\alpha}_{2,k}), \nonumber\\
&\lambda_{\bar{\delta}_{k}}^{(t+1)}=\lambda_{\bar{\delta}_{k}}^{(t)}+\frac{1}{\kappa^{(t)}}(\delta_{k}-\bar{\delta}_{k}), \lambda_{\hat{\delta}_{k}}^{(t+1)}=\lambda_{\hat{\delta}_{k}}^{(t)}+\frac{1}{\kappa^{(t)}}(\delta_{k}-\hat{\delta}_{k}),\lambda_{\tilde{\delta}_{k}}^{(t+1)}=\lambda_{\tilde{\delta}_{k}}^{(t)}+\frac{1}{\kappa^{(t)}}(\delta_{k}-\tilde{\delta}_{k}), \lambda_{\tilde{\Psi}_{1,k}}^{(t+1)}=\lambda_{\tilde{\Psi}_{1,k}}^{t}+\frac{1}{\kappa^{(t)}}(\Psi_{1,k}-\tilde{\Psi}_{1,k}), \nonumber\\
&\lambda_{\tilde{\Gamma}_{k}}^{(t+1)}=\lambda_{\tilde{\Gamma}_{k}}^{t}+\frac{1}{\kappa^{(t)}}(\Gamma_{k}-\tilde{\Gamma}_{k}),  \lambda_{\tilde{\nu}_{k}}^{(t+1)}=\lambda_{\tilde{\nu}_{k}}^{(t)}+\frac{1}{\kappa^{(t)}}(\nu_{k}-\tilde{\nu}_{k}), \lambda_{\tilde{u}_{k,k^{\prime}}}^{(t+1)}=\lambda_{\tilde{u}_{k,k^{\prime}}}^{(t)}+\frac{1}{\kappa^{(t)}}(\tilde{u}_{k,k^{\prime}}-\boldsymbol{f}_{k}^{H}\tilde{\boldsymbol{\mu}}_{k^{\prime}}), \nonumber\\
&\lambda_{\boldsymbol{\mu}_{k^{\prime}}}^{(t+1)}=\lambda_{\boldsymbol{\mu}_{k^{\prime}}}^{(t)}+\frac{1}{\kappa^{(t)}}(\boldsymbol{\mu}_{k^{\prime}}-\boldsymbol{B}_{k^{\prime}}\tilde{\boldsymbol{w}}_{k^{\prime}}),\lambda_{\tilde{\boldsymbol{\mu}}_{k^{\prime}}}^{(t+1)}=\lambda_{\tilde{\boldsymbol{\mu}}_{k^{\prime}}}^{(t)}+\frac{1}{\kappa^{(t)}}(\tilde{\boldsymbol{\mu}}_{k^{\prime}}-\boldsymbol{B}_{r}^{H}\boldsymbol{\Sigma}_{k}\boldsymbol{\mu}_{k^{\prime}}), \lambda_{\tilde{\boldsymbol{w}}_{k}}^{(t+1)}=\lambda_{\tilde{\boldsymbol{w}}_{k}}^{(t)}+\frac{1}{\kappa^{(t)}}(\tilde{\boldsymbol{w}}_{k}-\boldsymbol{w}_{k}), \nonumber\\
&\lambda_{\boldsymbol{f}_{k}}^{(t+1)}=\lambda_{\boldsymbol{f}_{k}}^{(t)}+\frac{1}{\kappa^{(t)}}(\tilde{\boldsymbol{f}}_{k}-\boldsymbol{f}_{k}), \lambda_{\boldsymbol{u}_{k}}^{(t+1)}=\lambda_{\boldsymbol{u}_{k}}^{(t)}+\frac{1}{\kappa^{(t)}}(\boldsymbol{u}_{k}-\boldsymbol{f}_{k}^{H}\boldsymbol{u}_{J}), \lambda_{\boldsymbol{u}_{J}}^{(t+1)}=\lambda_{\boldsymbol{u}_{J}}^{(t)}+\frac{1}{\kappa^{(t)}}(\boldsymbol{u}_{J}-\boldsymbol{B}_{J}\tilde{\boldsymbol{\Sigma}}\tilde{\boldsymbol{A}}_{J}\boldsymbol{z}),\nonumber\\
& \lambda_{\tilde{\boldsymbol{p}}_{k,n_{1}}}^{(t+1)}=\lambda_{\tilde{\boldsymbol{p}}_{k,n_{1}}}^{(t)}+\frac{1}{\kappa^{(t)}}(\tilde{\boldsymbol{p}}_{k,n_{1}}-\boldsymbol{p}_{k,n_{1}}-\boldsymbol{p}_{k,n_{2}}), \lambda_{\tilde{\boldsymbol{p}}_{r,m_{1}}}^{(t+1)}=\lambda_{\tilde{\boldsymbol{p}}_{r,m_{1}}}^{(t)}+\frac{1}{\kappa^{(t)}}(\tilde{\boldsymbol{p}}_{r,m_{1}}-\boldsymbol{p}_{r,m_{1}}-\boldsymbol{p}_{r,m_{2}}), \nonumber\\
&\boldsymbol{\lambda}_{\boldsymbol{B}_{k}}^{(t+1)}=\boldsymbol{\lambda}_{\boldsymbol{B}_{k}}^{(t)}+\frac{1}{\kappa^{(t)}}(\boldsymbol{A}(\boldsymbol{p}_{k})-\boldsymbol{B}_{k}), \boldsymbol{\lambda}_{\boldsymbol{B}_{r}}^{(t+1)}=\boldsymbol{\lambda}_{\boldsymbol{B}_{r}}^{(t)}+\frac{1}{\kappa^{(t)}}(\boldsymbol{A}(\boldsymbol{p}_{r})-\boldsymbol{B}_{r}), \boldsymbol{\lambda}_{\boldsymbol{B}_{J}}^{(t+1)}=\boldsymbol{\lambda}_{\boldsymbol{B}_{J}}^{(t)}+\frac{1}{\kappa^{(t)}}(\boldsymbol{A}_{J}(\boldsymbol{p}_{r})-\boldsymbol{B}_{J}),\label{pro19}
\end{align}
\hrulefill
\end{figure*}
where $t$ is the number of outer iterations. Following the approach outlined in \cite{9120361}, the flow of the proposed PDD-based algorithm is summarized in Fig.~\ref{FIGURE1}. The convergence analysis in \cite{9120361} confirms that the proposed PDD-based algorithm for joint hybrid beamforming and resource allocation converges to the set of stationary solutions for the problem (\ref{pro15}). Then, we provide a detailed description of the update procedure. According to the concept of the PDD algorithm in \cite{9120361},  we categorize the variable set into three blocks: \textbf{Block~1} comprises variables from set $\{\boldsymbol{w}_{k}$, $\tilde{\boldsymbol{f}}_{k}$,$\boldsymbol{B}_{k}$,$\boldsymbol{B}_{r}$,$\boldsymbol{B}_{J}$,$\gamma_{k}$,$\tilde{\gamma}_{k}$,$\tilde{\alpha}_{1,k}$,$\tilde{\alpha}_{2,k}$,$\bar{\delta}_{k}$,$\hat{\delta}_{k}$,$\tilde{\delta}_{k}$,$\tilde{\boldsymbol{\Psi}}_{1,k}$, $\tilde{\boldsymbol{\Gamma}}_{k}$,$\tilde{\boldsymbol{v}}_{k}$,$\tilde{\boldsymbol{p}}_{k,n_{1}}$,\\
$\tilde{\boldsymbol{p}}_{r,m_{1}}$, $\tilde{u}_{k,k^{\prime}}$,$\tilde{\nu}_{k}$,$\boldsymbol{u}_{k}\}$, \textbf{Block~2} includes variables from set $\{\gamma$,$\delta_{k}$,$\alpha_{1,k}$,$\alpha_{2,k}$,$\nu_{k}$,$\boldsymbol{\Psi}_{1,k}$,$\boldsymbol{f}_{k}$,$\boldsymbol{\mu}_{k^{\prime}}$,$\boldsymbol{\mu}_{k}$,$\boldsymbol{v}_{b,k}$, $\boldsymbol{p}_{k}$, $\boldsymbol{p}_{r}\}$, and \textbf{Block~3} consists of variables from set $\{\tilde{\boldsymbol{w}}_{k}$,$\boldsymbol{\mu}_{k^{\prime}}$,$\boldsymbol{u}_{J}\}$.

\begin{figure*}[htbp]
  \centering
  \includegraphics[width=0.9\textwidth, height=0.45\textwidth]{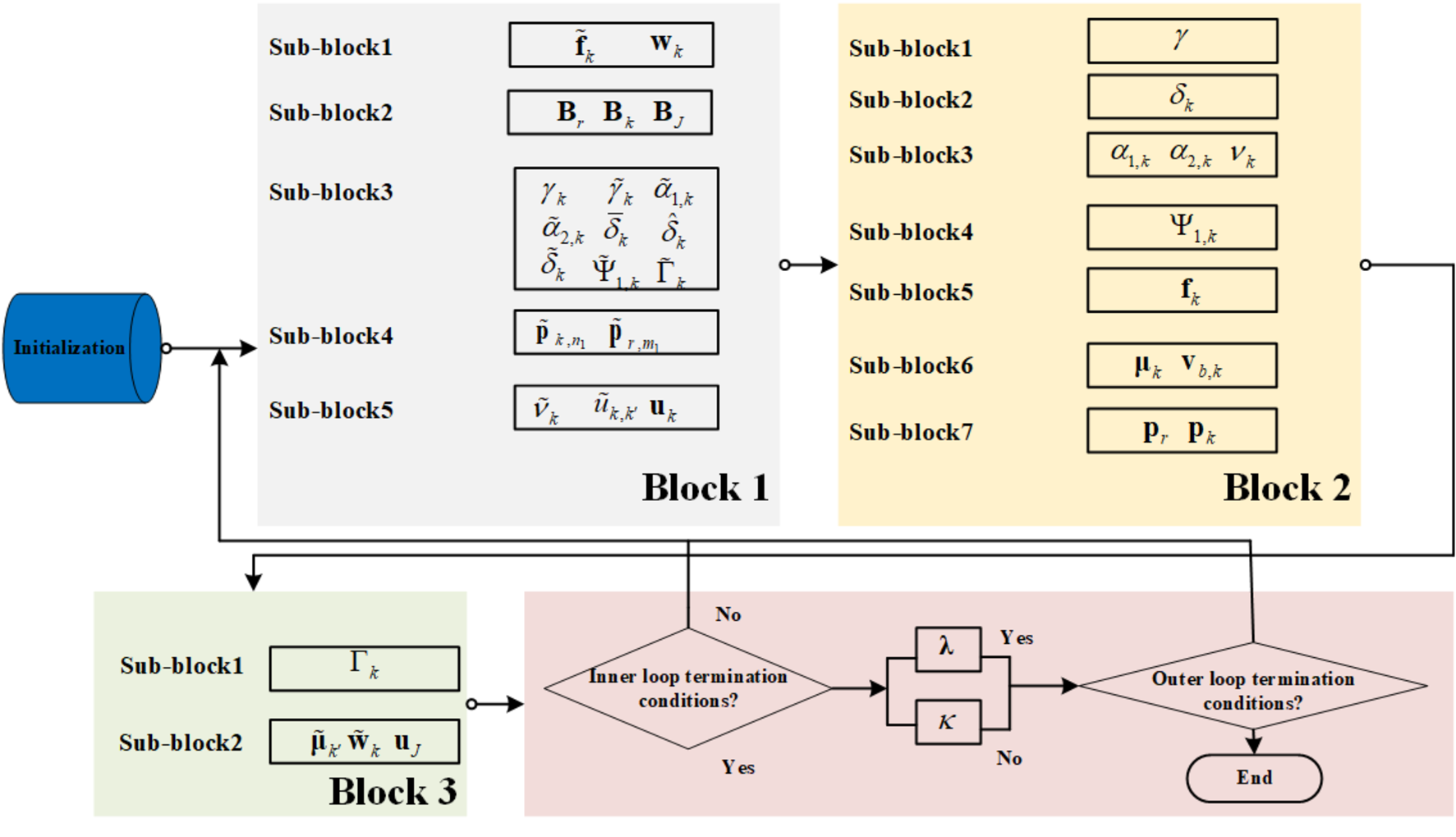}
  \captionsetup{justification=centering}
  \caption{Flowchart of the proposed SCA-based PDD algorithm.}
\label{FIGURE1}
\hrulefill
\end{figure*}


\subsection{Variables in \textbf{Block~1} update}
In this subsection, we divide the variables in \textbf{Block~1} into $5$ sub-blocks. \{$\boldsymbol{w}_{k}$, $\tilde{\boldsymbol{f}}_{k}$\} is called \textbf{sub-block~1}. \{$\boldsymbol{B}_{k}$, $\boldsymbol{B}_{r}$, $\boldsymbol{B}_{J}$\} is called \textbf{sub-block~2}. \{$\gamma_{k}$,$\tilde{\gamma}_{k}$,$\tilde{\alpha}_{1,k}$,$\tilde{\alpha}_{2,k}$,$\bar{\delta}_{k}$,$\hat{\delta}_{k}$,$\tilde{\delta}_{k}$,$\tilde{\Psi}_{1,k}$,$\tilde{\Gamma}_{k}$,$\tilde{\nu}_{k}$\} is called \textbf{sub-block~3}. \{$\tilde{\boldsymbol{p}}_{k,n_{1}}$,$\tilde{\boldsymbol{p}}_{r,m_{1}}$\} is called \textbf{sub-block~4}. \{$\tilde{u}_{k,k^{\prime}}$,$\tilde{\nu}_{k}$,$\boldsymbol{u}_{k}$\} is called \textbf{sub-block~5}. 

For \textbf{sub-block~1}, the subproblem with respect to $\boldsymbol{w}_{k}$ can be
solved with the aid of projection\cite{8606437}, with the solution given by
\begin{align}
\boldsymbol{w}_{k}=P_{k}\frac{\tilde{\boldsymbol{w}}_{k}-\kappa\boldsymbol{\lambda}_{\boldsymbol{w}}}{\|\tilde{\boldsymbol{w}}_{k}-\kappa\boldsymbol{\lambda}_{\boldsymbol{w}}\|+\max(0,P_{k}-\|\tilde{\boldsymbol{w}}_{k}-\kappa\boldsymbol{\lambda}_{\boldsymbol{w}}\|)}.\label{pro22}
\end{align}
Similarly, $\tilde{\boldsymbol{f}}_{k}$ can be obtained using the same method.  For \textbf{sub-block~2}, the subproblem with respect to $\boldsymbol{B}_{k^{\prime}}$ only appears in the summand of the 12-th and the 20-th terms of the objective of (\ref{pro15}). Thus, the second subproblem is given by
\begin{align}
\min_{|\boldsymbol{B}_{k^{\prime}}(l,n_{1})|=1}&\sum\nolimits_{k^{\prime}=1}^{K}\|\boldsymbol{\mu}_{k^{\prime}}-\boldsymbol{B}_{k^{\prime}}\tilde{\boldsymbol{w}}_{k^{\prime}}+\kappa\boldsymbol{\lambda}_{\boldsymbol{\mu}_{k^{\prime}}}\|^{2}&\nonumber\\
&+\|\boldsymbol{A}(\boldsymbol{p}_{k^{\prime}})-\boldsymbol{B}_{k^{\prime}}+\kappa\boldsymbol{\lambda}_{\boldsymbol{B}_{k^{\prime}}}\|^{2}.&\label{pro23}
\end{align}
With proper rearrangement, the problem can be equivalently reformulated as follows
\begin{align}
\min_{|\boldsymbol{B}_{k^{\prime}}(l,n_{1})|=1}&\sum\nolimits_{k^{\prime}=1}^{K}\mathrm{tr}(\boldsymbol{B}_{k^{\prime}}^{H}\boldsymbol{B}_{k^{\prime}}(
\tilde{\boldsymbol{w}}_{k^{\prime}}\tilde{\boldsymbol{w}}_{k^{\prime}}^{H}+\boldsymbol{I}))-2\mathrm{Re}\{\mathrm{tr}&\nonumber\\
&(\boldsymbol{B}_{k^{\prime}}^{H}(\boldsymbol{\mu}_{k^{\prime}}+\kappa\boldsymbol{\lambda}_{\boldsymbol{\mu}_{k^{\prime}}}+\boldsymbol{A}(\boldsymbol{p}_{k^{\prime}})+\kappa\boldsymbol{\lambda}_{\boldsymbol{B}_{k^{\prime}}}))\}.&\label{pro24}
\end{align}
Given that the unit modulus constraints are separable, a one-iteration block coordinate descent (BCD) type algorithm can be used to solve the problem (\ref{pro24}) iteratively. Specifically, problem (\ref{pro24}) is rewritten as
\begin{align}
\min_{\mathcal{M}}&f(\boldsymbol{B}_{k^{\prime}})=\mathrm{tr}(\boldsymbol{B}_{k^{\prime}}^{H}\boldsymbol{B}_{k^{\prime}}(
\tilde{\boldsymbol{w}}_{k^{\prime}}\tilde{\boldsymbol{w}}_{k^{\prime}}^{H}+\boldsymbol{I}))-2\mathrm{Re}\{\mathrm{tr}(\boldsymbol{B}_{k^{\prime}}^{H}&\nonumber\\
&((\boldsymbol{\mu}_{k^{\prime}}+\kappa\boldsymbol{\lambda}_{\boldsymbol{\mu}_{k^{\prime}}})\tilde{\boldsymbol{w}}_{k^{\prime}}^{H}+\boldsymbol{A}(\boldsymbol{p}_{k^{\prime}})+\kappa\boldsymbol{\lambda}_{\boldsymbol{B}_{k^{\prime}}}))\}.&\label{pro_25}
\end{align}   
Since the unit modulus constraints are separable, we can use the one-iteration BCD-type algorithm presented in
the Appendix of \cite{8332507} to solve problem (\ref{pro_25}).
For \textbf{sub-block~3}, the subproblem with respect to $\chi_{1}=\{\gamma_{k},\tilde{\gamma}_{k},\tilde{\alpha}_{1,k},\tilde{\alpha}_{2,k},\bar{\delta}_{k},\hat{\delta}_{k},\tilde{\delta}_{k},\tilde{\Psi}_{1,k},\tilde{\Gamma}_{k}\}$ is given by
\begin{subequations}
\begin{align}
\min_{\chi_{1}}&~|\gamma-\gamma_{k}+\kappa\lambda_{\gamma_{k}}|^{2}+|\gamma-\tilde{\gamma}_{k}+\kappa\lambda_{\tilde{\gamma}_{k}}|^{2}+|\gamma-\gamma_{k}+\kappa\lambda_{\gamma_{k}}|^{2}&\nonumber\\
&+|\alpha_{1,k}-\tilde{\alpha}_{1,k}+\kappa\lambda_{\tilde{\alpha}_{1,k}}|^{2}+|\alpha_{2,k}-\tilde{\alpha}_{2,k}+\kappa\lambda_{\tilde{\alpha}_{2,k}}|^{2}+&\nonumber\\
&|\delta_{k}-\bar{\delta}_{k}+\kappa\lambda_{\bar{\delta}_{k}}|^{2}+|\delta_{k}-\hat{\delta}_{k}+\kappa\lambda_{\hat{\delta}_{k}}|^{2}+&\nonumber\\
&|\delta_{k}-\tilde{\delta}_{k}+\kappa\lambda_{\tilde{\delta}_{k}}|^{2}+|\Psi_{1,k}-\tilde{\Psi}_{1,k}+\kappa\lambda_{\Psi_{1,k}}|^{2}+&\nonumber\\
&|\Gamma_{k}-\tilde{\Gamma}_{k}+\kappa\lambda_{\Gamma_{k}}|^{2}+|\nu_{k}-\tilde{\nu}_{k}+\kappa\lambda_{\nu_{k}}|^{2},&\label{pro25a}\\
\mbox{s.t.}~
&(\ref{pro16}),(\ref{pro14d}),(\ref{pro14g}).&\label{pro25b}
\end{align}\label{pro25}%
\end{subequations}
Note that there exists only one constraint in this subproblem,
 so it can be solved in closed form through the
Lagrange multiplier method. The corresponding Lagrange
function can be written as (\ref{pro26}) at the top of this page,
\begin{figure*}
\begin{align}
&\mathcal{L}_{1}(\chi_{1})=|\gamma-\gamma_{k}+\kappa\lambda_{\gamma_{k}}|^{2}+|\gamma-\tilde{\gamma}_{k}+\kappa\lambda_{\tilde{\gamma}_{k}}|^{2}+|\gamma-\gamma_{k}+\kappa\lambda_{\gamma_{k}}|^{2}+|\alpha_{1,k}-\tilde{\alpha}_{1,k}+\kappa\lambda_{\tilde{\alpha}_{1,k}}|^{2}+|\alpha_{2,k}-\tilde{\alpha}_{2,k}+\kappa\lambda_{\tilde{\alpha}_{2,k}}|^{2}+\nonumber\\
&|\delta_{k}-\bar{\delta}_{k}+\kappa\lambda_{\bar{\delta}_{k}}|^{2}+|\delta_{k}-\hat{\delta}_{k}+\kappa\lambda_{\hat{\delta}_{k}}|^{2}+|\delta_{k}-\tilde{\delta}_{k}+\kappa\lambda_{\tilde{\delta}_{k}}|^{2}+|\Psi_{1,k}-\tilde{\Psi}_{1,k}+\kappa\lambda_{\Psi_{1,k}}|^{2}+|\Gamma_{k}-\tilde{\Gamma}_{k}+\kappa\lambda_{\Gamma_{k}}|^{2}+\nonumber\\
&|\nu_{k}-\tilde{\nu}_{k}+\kappa\lambda_{\nu_{k}}|^{2}+\kappa_{1,k}(\tilde{\alpha}_{1,k}+\tilde{\alpha}_{2,k}-\gamma_{k})+\kappa_{2,k}(\bar{\delta}_{k}^{2}+\frac{\Delta_{k}^{2}}{\tilde{\Psi}_{1,k}^{2}}-2\alpha_{1,k})+\kappa_{3,k}(\tilde{\delta}_{k}^{2}+\frac{\Delta_{k}^{2}}{B^{2}\Gamma_{k}^{2}}-2\alpha_{2,k})+\kappa_{4,k}((1-\hat{\delta}_{k})^{2}\nonumber\\
&+\frac{\Delta_{k}^{2}}{\Psi_{2,k}^{2}}-\tilde{\gamma}_{k})+\kappa_{5,k}(\tilde{\Gamma}_{k}-\log_{2}(1+\nu_{k}^{(t-1)})-(\nu_{k}-\nu_{k}^{(t-1)})/(2\ln2(1+\nu_{k}^{(t-1)}))),\label{pro26}
\end{align}
\hrulefill
\end{figure*}
where $\kappa_{1,k},\kappa_{2,k},\kappa_{3,k},\kappa_{4,k}\geq 0$ denotes the Lagrange multiplier  for constraints (\ref{pro16}), (\ref{pro14d}), (\ref{pro14g}). By computating the ﬁrst order optimality
condition of $\mathcal{L}_{1}(\chi_{1})$, the optimal value of
 can be derived as (\ref{pro27}) at this page.
\begin{figure*}
\begin{align}
&\gamma_{k}(\kappa_{1,k})=\frac{2(\gamma+\kappa\lambda_{\gamma_{k}})-\kappa_{1,k}}{2}, \tilde{\gamma}_{k}(\kappa_{4,k})=\frac{2(\gamma+\kappa\lambda_{\tilde{\gamma}_{k}})-\kappa_{1,k}}{2},
 \tilde{\alpha}_{1,k}(\kappa_{1,k})=\frac{2(\alpha_{1,k}+\kappa\lambda_{\alpha_{1,k}})-\kappa_{1,k}}{2}, \nonumber\\
&\tilde{\alpha}_{2,k}(\kappa_{1,k})=\frac{2(\alpha_{2,k}+\kappa\lambda_{\alpha_{2,k}})-\kappa_{1,k}}{2}, \tilde{\Psi}_{1,k}(\kappa_{2,k})=\frac{2(\Psi_{1,k}+\kappa\lambda_{\Psi_{1,k}})-2\kappa_{2,k}\Delta_{k}^{2}/(\tilde{\Psi}_{1,k}^{(t-1)})^{3}}{2}, 
\bar{\delta}_{k}(\kappa_{2,k})=\frac{\delta_{k}+\kappa\lambda_{\bar{\delta}_{k}}}{1+\kappa_{2,k}},\nonumber\\
&\tilde{\delta}_{k}(\kappa_{3,k})=\frac{\delta_{k}+\kappa\lambda_{\tilde{\delta}_{k}}}{1+\kappa_{3,k}}, 
\hat{\delta}_{k}(\kappa_{4,k})=\frac{\delta_{k}+\kappa\lambda_{\hat{\delta}_{k}}-\kappa_{4,k}}{1-\kappa_{4,k}}.\label{pro27}
\end{align}    
\hrulefill
\end{figure*}
Let us denote the optimal $\kappa_{1,k}$, $\kappa_{2,k}$, $\kappa_{3,k}$, $\kappa_{4,k}$ as $\kappa_{1,k}^{*}$, $\kappa_{2,k}^{*}$, $\kappa_{3,k}^{*}$, $\kappa_{4,k}^{*}$, then $\kappa_{1,k}^{*}$, $\kappa_{2,k}^{*}$, $\kappa_{3,k}^{*}$, $\kappa_{4,k}^{*}$ are determined
 to fulfill the complementary slackness condition
of (\ref{pro25b}), and can be given in (\ref{pro28}) at the top of the next page.
\begin{figure*}
\begin{align}
&\kappa_{1,k}^{*}=\max\left\{0,\frac{2(\alpha_{1,k}+\kappa\lambda_{\alpha_{1,k}})+2(\alpha_{2,k}+\kappa\lambda_{\alpha_{2,k}})-2(v+\kappa\lambda_{\gamma_{k}})}{3}\right\},
\kappa_{2,k}^{*}=\max\left\{0,\pm\frac{\delta_{k}+\kappa\lambda_{\bar{\delta}_{k}}}{\sqrt{2\alpha_{1,k}-\Delta_{k}^{2}/\tilde{\Psi}_{1,k}^{2}}}-1\right\},\nonumber\\
&\kappa_{3,k}^{*}=\max\left\{0,\pm\frac{\delta_{k}+\kappa\lambda_{\tilde{\delta}_{k}}}{\sqrt{2\alpha_{2,k}-\Delta_{k}^{2}/B^{2}\Gamma_{k}^{2}}}-1\right\},
\kappa_{4,k}^{*}=\max\left\{0,1\pm\frac{\delta_{k}-\kappa\lambda_{\hat{\delta}_{k}}-1}{\sqrt{\tilde{\gamma}_{k}-\Delta_{k}^{2}/\Psi_{2,k}^{2}}}\right\}.\label{pro28}
\end{align}    
\hrulefill
\end{figure*}
For \textbf{sub-block~4}, similarly, Lagrange multiplier for the subproblem with respect to $\tilde{\boldsymbol{p}}_{k,n_{1}}$ is given in (\ref{pro29}) at the top of this page.
\begin{figure*}
\begin{align}
&\mathcal{L}(\tilde{\boldsymbol{p}}_{k,n_{1}})=|\tilde{\boldsymbol{p}}_{k,n_{1}}-(\boldsymbol{p}_{k,n_{1}}-\boldsymbol{p}_{k,n_{2}})+\boldsymbol{\lambda}_{\tilde{\boldsymbol{p}}_{k,n_{1}}}|^{2}+\kappa_{5,k}(d-\|\tilde{\boldsymbol{p}}_{k,n_{1}}^{(t-1)}\|_{2}+2\tilde{\boldsymbol{p}}_{k,n_{1}}^{(t-1)}\tilde{\boldsymbol{p}}_{k,n_{1}}^{T})&.\label{pro29}
\end{align}
\hrulefill
\end{figure*}
$\tilde{\boldsymbol{p}}_{k,n_{1}}$ is given by
\begin{align}
&\tilde{\boldsymbol{p}}_{k,n_{1}}(\kappa_{5,k})=(\boldsymbol{p}_{k,n_{1}}-\boldsymbol{p}_{k,n_{2}})-\boldsymbol{\lambda}_{\tilde{\boldsymbol{p}}_{k,n_{1}}}+\kappa_{5,k}\tilde{\boldsymbol{p}}_{k,n_{1}}^{(t-1)}.\label{pro_29}
\end{align}
The optimal Lagrange multiplier $\kappa_{5,k}^{*}$ is given in (\ref{pro31}). Similarly, 
$\tilde{\boldsymbol{p}}_{r,m_{1}}$ can be obtained using the same method.
\begin{figure*}[hbt]
\begin{align}
&\kappa_{5,k}^{*}=\max\left\{0, (\|\tilde{\boldsymbol{p}}_{k,n_{1}}^{(t-1)}\|_{2}-d-2(\tilde{\boldsymbol{p}}_{k,n_{1}}^{(t-1)})^{T}((\boldsymbol{p}_{k,n_{1}}-\boldsymbol{p}_{k,n_{2}})-\boldsymbol{\lambda}_{\tilde{\boldsymbol{p}}_{k,n_{1}}}))/(2\tilde{\boldsymbol{p}}_{k,n_{1}}^{(t-1)})^{T}\tilde{\boldsymbol{p}}_{k,n_{1}}^{(t-1)}\right\}. \label{pro31}
\end{align}
\hrulefill
\end{figure*}
For \textbf{sub-block~5}, the subproblem with respect to $\chi_{3}=\{\tilde{u}_{k,k^{\prime}},\tilde{\nu}_{k},\boldsymbol{u}_{k}\}$ is given by
\begin{subequations}
\begin{align}
\min_{\chi_{3}}&~\sum_{k^{\prime}=1}^{K}|\tilde{u}_{k,k^{\prime}}-\boldsymbol{u}_{k}\tilde{\boldsymbol{\mu}}_{k^{\prime}}+\kappa\lambda_{\tilde{u}_{k,k^{\prime}}}|^{2}+|\nu_{k}-\tilde{\nu}_{k}+\kappa\lambda_{\tilde{\nu}_{k}}|^{2}+\nonumber\\
&\|\boldsymbol{u}_{k}-\boldsymbol{f}_{k}^{H}\boldsymbol{u}_{J}+\kappa\boldsymbol{\lambda}_{\boldsymbol{u}_{k}}\|^{2},&\label{pro32a}\\
\mbox{s.t.}~
&(\ref{pro17}).&\label{pro32b}
\end{align}\label{pro32}%
\end{subequations}
Similarly, by introducing the Lagrange multiplier $\kappa_{10,k}\geq 0$ to constraint (\ref{pro32b}), the optimal $\chi_{3}$ can be expressed as
\begin{align}
&\tilde{\nu}_{k}(\kappa_{6,k})=\frac{2(\nu_{k}+\kappa\lambda_{\nu_{k}})-\kappa_{6,k}|\tilde{u}_{k,k}^{(t-1)}|^{2}/(\tilde{\nu}_{k}^{(t-1)})^{2}}{2},\nonumber\\
&\boldsymbol{u}_{k}(\kappa_{6,k})=\boldsymbol{\Phi}^{-1}(\kappa_{6,k})\boldsymbol{\phi}(\kappa_{6,k}),\nonumber\\
&\tilde{u}_{k,k^{\prime}}(\kappa_{6,k})=\frac{\boldsymbol{u}_{k}\tilde{\boldsymbol{\mu}}_{k^{\prime}}-\kappa\lambda_{\tilde{u}_{k,k^{\prime}}}}{1+\kappa_{6,k}},~\forall~k\neq k^{\prime},\nonumber\\
&\tilde{u}_{k,k^{\prime}}(\kappa_{6,k})=\boldsymbol{u}_{k}\tilde{\boldsymbol{\mu}}_{k,k}-\kappa\lambda_{\tilde{u}_{k,k}}+\kappa_{6,k}\frac{\tilde{u}_{k,k}^{(t-1)}}{\tilde{\nu}_{k}^{(t-1)}},\label{pro33}
\end{align}
where $\boldsymbol{\Phi}(\kappa_{6,k})=(1+\kappa_{6,k}\sigma_{r}^{2})\boldsymbol{I}+\kappa_{6,k}\frac{\tilde{\boldsymbol{\Phi}}_{k}}{1+\kappa_{6,k}}$, and $\boldsymbol{\phi}(\kappa_{6,k})=\tilde{\boldsymbol{v}}_{b,k}-\kappa\lambda_{\boldsymbol{u}_{k}}+\kappa_{6,k}\frac{\sum_{k\neq k^{\prime}}\tilde{\boldsymbol{\mu}}_{k^{\prime}}\kappa\lambda_{\tilde{u}_{k,k^{\prime}}}}{1+\kappa_{6,k}}+\kappa_{6,k}\tilde{\boldsymbol{\mu}}_{k,k}\tilde{u}_{k,k}^{(t-1)}/\tilde{\nu}_{k}^{(t-1)}$. We also let $\tilde{\boldsymbol{\Phi}}_{k}=\sum_{k\neq k^{\prime}}\tilde{\boldsymbol{\mu}}_{k^{\prime}}\tilde{\boldsymbol{\mu}}_{k^{\prime}}^{H}$. The optimal multiplier
$\kappa_{6,k}$ should be determined such that the complementarity slackness condition is satisﬁed. Let us deﬁne 
\begin{align}
&Q_{k}(\chi_{3})=\sum_{k^{\prime}\neq k}|\tilde{u}_{k,k^{\prime}}|^{2}+\|\tilde{\boldsymbol{f}}_{k}\|^{2}\sigma^{2}_{r}+\|\boldsymbol{u}_{J}\|^{2}-(\tilde{u}_{k,k}^{(t-1),*}\nonumber\\
&\tilde{u}_{k,k}/\tilde{\nu}_{k}^{(t-1)}+\tilde{u}_{k,k}^{*}\tilde{u}_{k,k}^{(t-1)}/\tilde{\nu}_{k}^{(t-1)}-|\tilde{u}_{k,k}^{(t-1)}|^{2}\tilde{\nu}_{k}/(\tilde{\nu}_{k}^{(t-1)})^{2})\nonumber\\
&\leq 0.\label{pro34}
\end{align}
When $Q_{k}(\chi_{3}(0))\leq 0$, we have
the optimal $\tilde{u}_{k,k^{\prime}}=\tilde{u}_{k,k^{\prime}}(0)$,$\tilde{\nu}_{k}=\tilde{\nu}_{k}(0)$, and $\boldsymbol{u}_{k}=\boldsymbol{u}_{k}(0)$, otherwise we must have $Q_{k}(\chi_{3})=0$, which is equivalent to
\begin{align}
&\boldsymbol{u}_{k}^{H}(\kappa_{6,k})\hat{\boldsymbol{\Phi}}(\kappa_{6,k})\boldsymbol{u}_{k}(\kappa_{6,k})-\boldsymbol{u}_{k}^{H}(\kappa_{6,k})\boldsymbol{m}(\kappa_{6,k})-\nonumber\\
&\boldsymbol{m}(\kappa_{6,k})^{H}\boldsymbol{u}_{k}(\kappa_{6,k})+\varpi_{k}(\kappa_{6,k})=0,\label{pro35}
\end{align}
where $\hat{\boldsymbol{\Phi}}(\kappa_{6,k})=\sigma_{r}^{2}\boldsymbol{I}+\frac{\tilde{\boldsymbol{\Phi}}_{k}}{(1+\kappa_{6,k})^{2}}$, $\boldsymbol{m}(\kappa_{6,k})$=$\frac{\sum_{k^{\prime}\neq k}\kappa\lambda_{\tilde{u}_{k,k^{\prime}}}^{(t-1)}\tilde{\boldsymbol{\mu}}_{k^{\prime}}}{(1+\kappa_{6,k})^{2}}$+$\frac{\tilde{u}_{k,k}^{(t-1)}}{\tilde{\nu}_{k}^{(t-1)}}\tilde{\boldsymbol{\mu}}_{k^{\prime}}$, and $\varpi_{k}(\kappa_{6,k})$=$\frac{\sum\nolimits_{k^{\prime}\neq k}\kappa^{2}|\lambda_{\tilde{u}_{k^{\prime},k}}|^{2}}{(1+\kappa_{6,k})^{2}}$+$\frac{|\tilde{u}_{k,k}^{(t-1)}|^{2}}{(\tilde{\nu}_{k}^{(t-1)})^{2}}$+$\frac{2(\nu_{k}+\kappa\lambda_{\nu_{k}})}{2}$-$\frac{\kappa_{6,k}|\tilde{u}_{k,k}^{(t-1)}|^{2}/(\tilde{\nu}_{k}^{(t-1)})^{2}}{2}$+$2\mathrm{Re}\{\frac{\tilde{u}_{k,k}^{(t-1)}}{\tilde{\nu}_{k}^{(t-1)}}(\kappa\lambda_{\tilde{u}_{k,k}}+\kappa_{6,k}\frac{\tilde{u}_{k,k}^{(t-1)}}{\tilde{\nu}_{k}^{(t-1)}})\}$. For the Hermitian matrix $\tilde{\boldsymbol{\Phi}}_{k}$, we put forward the following
decomposition:
\begin{align}
\tilde{\boldsymbol{\Phi}}_{k}=\tilde{\boldsymbol{V}}_{k}\boldsymbol{\Lambda}_{k}\tilde{\boldsymbol{V}}_{k}^{H},\label{pro36}
\end{align}
in which $\tilde{\boldsymbol{V}}_{k}$ denotes a unitary matrix which contains the eigenvectors of $\tilde{\boldsymbol{\Phi}}_{k}$, and $\boldsymbol{\Lambda}_{k}$ is a diagonal matrix consisting of the eigenvectors of $\tilde{\boldsymbol{\Phi}}_{k}$. Then we have
\begin{align}
\boldsymbol{\Phi}^{-1}(\kappa_{6,k})=\tilde{\boldsymbol{V}}_{k}((1+\kappa_{6,k}\sigma_{r}^{2})\boldsymbol{I}+\boldsymbol{\Lambda}_{k}\frac{\kappa_{6,k}}{1+\kappa_{6,k}})^{-1}\tilde{\boldsymbol{V}}_{k}^{H}.\label{pro37}
\end{align}
Similarly, $\hat{\boldsymbol{\Phi}}(\kappa_{6,k})$ can be rewritten as
\begin{align}
\hat{\boldsymbol{\Phi}}(\kappa_{6,k})=\tilde{\boldsymbol{V}}_{k}(\sigma_{r}^{2}\boldsymbol{I}+\frac{\boldsymbol{\Lambda}_{k}}{(1+\kappa_{6,k})^{2}})\tilde{\boldsymbol{V}}_{k}^{H}.\label{pro38}
\end{align}
Let $\check{\boldsymbol{\Phi}}(\kappa_{6,k})=(1+\kappa_{6,k}\sigma_{r}^{2})\boldsymbol{I}+\boldsymbol{\Lambda}_{k}\frac{\kappa_{6,k}}{1+\kappa_{6,k}}$ and $\ddot{\boldsymbol{\Phi}}(\kappa_{6,k})=\sigma_{r}^{2}\boldsymbol{I}+\frac{\boldsymbol{\Lambda}_{k}}{(1+\kappa_{6,k})^{2}}$. Then (\ref{pro34}) can be rewritten
as
\begin{align}
&\mathrm{tr}(\check{\boldsymbol{\Phi}}^{-1}\ddot{\boldsymbol{\Phi}}\check{\boldsymbol{\Phi}}^{-1}\tilde{\boldsymbol{V}}_{k}^{H}\boldsymbol{\phi}\boldsymbol{\phi}^{H}\tilde{\boldsymbol{V}}_{k})+\mathrm{tr}(\check{\boldsymbol{\Phi}}^{-1}\tilde{\boldsymbol{V}}_{k}^{H}(\boldsymbol{\phi}\ddot{\boldsymbol{\Phi}}^{H}+\boldsymbol{\phi}^{H}\ddot{\boldsymbol{\Phi}})\tilde{\boldsymbol{V}}_{k})\nonumber\\
&+t_{k}=0.\label{pro39}
\end{align}
Since $\check{\boldsymbol{\Phi}}$ and $\ddot{\boldsymbol{\Phi}}$ are both diagonal matrices, (\ref{pro39}) can be
easily solved using one dimensional search. Finally, by
substituting the optimal $\kappa_{6,k}$, we obtain the solution for
$\chi_{3}$.

\subsection{Variables in \textbf{Block~2} update}
In this subsection, we divide the variables in \textbf{Block~2} into $7$ sub-blocks. $\gamma$ is called \textbf{sub-block~1}. \{$\delta_{k}$\} is called \textbf{sub-block~2}. \{$\alpha_{1,k}$,$\alpha_{2,k}$,$\nu_{k}$\} is called \textbf{sub-block~3}. \{$\boldsymbol{\Psi}_{1,k}$\} is called \textbf{sub-block~4}. \{$\boldsymbol{f}_{k}$, $\boldsymbol{\mu}_{k^{\prime}}$\} is called \textbf{sub-block~5}. \{$\boldsymbol{\mu}_{k}$, $\boldsymbol{v}_{b,k}$\} is called \textbf{sub-block~6}. \{$\boldsymbol{p}_{k}$, $\boldsymbol{p}_{r}$\} is called \textbf{sub-block~7}.

For the \textbf{sub-block~1}, the optimal solution of the subproblem with respect to $\gamma$ is given by
\begin{align}
\gamma=\sum\nolimits_{k=1}^{K}(\gamma_{k}-\kappa\lambda_{\gamma_{k}}+\tilde{\gamma}_{k}-\kappa\lambda_{\tilde{\gamma}_{k}})/2.\label{pro41}
\end{align}
For the \textbf{sub-block~2}, the optimal solution of the subproblem with respect to $\delta_{k}$ is given in (\ref{pro42}) at the top of this page, where function $\epsilon(\delta_{k})$ denotes as $\epsilon(\delta_{k})=|\delta_{k}-\bar{\delta}_{k}+\kappa\lambda_{\bar{\delta}_{k}}|^{2}+|\delta_{k}-\hat{\delta}_{k}+\kappa\lambda_{\hat{\delta}_{k}}|^{2}+|\delta_{k}-\tilde{\delta}_{k}+\kappa\lambda_{\tilde{\delta}_{k}}|^{2}$.
\begin{figure*}
\begin{align}
\delta_{k}=\left\{\begin{matrix}
(\bar{\delta}_{k}+\hat{\delta}_{k}+\tilde{\delta}_{k}-2\kappa(\lambda_{\bar{\delta}_{k}}+\lambda_{\hat{\delta}_{k}}+\lambda_{\tilde{\delta}_{k}})/3,&~\textrm{if}~ 0\leq (\bar{\delta}_{k}+\hat{\delta}_{k}+\tilde{\delta}_{k}-2\kappa(\lambda_{\bar{\delta}_{k}}+\lambda_{\hat{\delta}_{k}}+\lambda_{\tilde{\delta}_{k}})/3\leq 1\\
1, &~\textrm{else}~\textrm{if}~\pi(1)<\pi(0)\\
0, &~\textrm{else}\\
\end{matrix}\right.\label{pro42}
\end{align}
\hrulefill
\end{figure*}
For the \textbf{sub-block~3}, the subproblem with respect to $\chi_{4}=\{\alpha_{1,k},\alpha_{2,k},\nu_{k}\}$ is given by
\begin{subequations}
\begin{align}
\min_{\chi_{4}}&~|\alpha_{1,k}-\tilde{\alpha}_{1,k}+\kappa\lambda_{\alpha_{1,k}}|^{2}+|\alpha_{2,k}-\tilde{\alpha}_{2,k}+\kappa\lambda_{\alpha_{2,k}}|^{2}+&\nonumber\\
&|\nu_{k}-\tilde{\nu}_{k}+\kappa\lambda_{\tilde{\nu}_{k}}|^{2},&\label{pro43a}\\
\mbox{s.t.}~
&(\ref{pro16}).&
\end{align}
\end{subequations}
Similarly, we use Lagrange multiplier method to obtain the solution of  $\chi_{4}$ and $\alpha_{1,k}$,$\alpha_{2,k}$,$\nu_{k}$ are denoted by
\begin{align}
&\alpha_{1,k}(\kappa_{7,k})=\tilde{\alpha}_{1,k}+\kappa_{7,k}-\kappa\lambda_{\alpha_{1,k}},\nonumber\\
&\alpha_{2,k}(\kappa_{8,k})=\tilde{\alpha}_{2,k}+\kappa_{8,k}-\kappa\lambda_{\alpha_{2,k}},\nonumber\\
&\nu_{k}(\kappa_{9,k})=\frac{2(\tilde{\nu}_{k}-\kappa\lambda_{\nu_{k}})+\kappa_{9,k}\frac{1}{2\ln 2(1+\nu_{k}^{(t-1)})}}{2}.\label{pro44}
\end{align}
The Lagrange multipliers are expressed as
\begin{align}
&\kappa_{7,k}^{*}=\max\left\{0,\frac{\bar{\delta}_{k}^{2}}{2}+\frac{\Delta_{k}^{2}}{2\tilde{\Psi}_{1,k}^{2}}-(\tilde{\alpha}_{1,k}-\kappa\lambda_{\alpha_{1,k})}\right\},\nonumber\\
&\kappa_{8,k}^{*}=\max\left\{0,\frac{\tilde{\delta}_{k}^{2}}{2}+\frac{\Delta_{k}^{2}}{2B^{2}\Gamma_{k}^{2}}-(\tilde{\alpha}_{2,k}-\kappa\lambda_{\alpha_{2,k})}\right\},\nonumber\\
&\kappa_{9,k}^{*}=\max\left\{0,\Gamma_{k}+\kappa\lambda_{\Gamma_{k}}+\frac{\nu_{k}+\kappa\lambda_{\nu_{k}}}{2\ln 2(1+\nu_{k}^{(t-1)})}-\right.\nonumber\\
&\left.\log_{2}(1+\nu_{k}^{(t-1)})-\frac{\nu_{k}^{(t-1)}}{2\ln 2(1+\nu_{k}^{(t-1)})}\right\}.\label{pro45}
\end{align}
For the \textbf{sub-block~4}, 
the subproblem with respect to $\Psi_{1,k}$ is expressed as
\begin{subequations}
\begin{align}
\min_{\Psi_{1,k}}&~|\Psi_{1,k}-\tilde{\Psi}_{1,k}+\kappa\lambda_{\Psi_{1,k}}|^{2},&\\
\mbox{s.t.}~
&(\ref{pro14g}).&\label{pro_46}
\end{align}
\end{subequations}
By using the Lagrange multiplier method, $\Psi_{1,k}$ and the Lagrange multiplier $\kappa_{10,k}$ are given by
\begin{align}
&\kappa_{10,k}=\max\left\{0,\sum\nolimits_{k=1}^{k}2((\tilde{\Psi}_{1,k}-\kappa\lambda_{\Psi_{1,k}})-\Psi_{max})/K\right\},\nonumber\\
&\Psi_{1,k}=(2(\tilde{\Psi}_{1,k}-\kappa\lambda_{\Psi_{1,k}})-\kappa_{10,k})/2. \label{pro__46}
\end{align}
For the \textbf{sub-block~5}, the subproblem with regarding to $\boldsymbol{f}_{k}$ is given by
\begin{align}
\min_{\boldsymbol{f}_{k}}\|\boldsymbol{f}_{k}-\tilde{\boldsymbol{f}}_{k}+\kappa\boldsymbol{\lambda}_{\boldsymbol{f}_{k}}\|^{2}+\|\boldsymbol{u}_{k}-\boldsymbol{f}_{k}^{H}\boldsymbol{u}_{J}+\kappa\boldsymbol{\lambda}_{\boldsymbol{u}_{k}}\|^{2}. \label{pro47}
\end{align}
The optimal solution of $\boldsymbol{f}_{k}$ is denoted as
\begin{align}
\boldsymbol{f}_{k}=(\tilde{\boldsymbol{f}}_{k}-\kappa\boldsymbol{\lambda}_{\boldsymbol{f}_{k}}+(\boldsymbol{u}_{k}+\kappa\boldsymbol{\lambda}_{\boldsymbol{u}_{k}})\boldsymbol{u}_{J}^{H})(\boldsymbol{I}+\boldsymbol{u}_{J}^{*}\boldsymbol{u}_{J}^{T})^{-1}.\label{pro48}
\end{align}
For the \textbf{sub-block~6}, the subproblem with respect to $\chi_{5}=\{\boldsymbol{\mu}_{k},\boldsymbol{v}_{b,k}\}$ is given by
\begin{align}
\min_{\chi_{5}}&~\|\boldsymbol{\mu}_{k^{\prime}}-\boldsymbol{B}_{t,k^{\prime}}\tilde{\boldsymbol{w}}_{k^{\prime}}+\kappa\boldsymbol{\lambda}_{\boldsymbol{\mu}_{k^{\prime}}}\|^{2}+\|\tilde{\boldsymbol{\mu}}_{k^{\prime}}-\boldsymbol{B}_{r}^{H}\boldsymbol{\Sigma}_{k}\boldsymbol{\mu}_{k^{\prime}}&\nonumber\\
&+\kappa\boldsymbol{\lambda}_{\tilde{\boldsymbol{\mu}}_{k^{\prime}}}\|^{2}
+\|\boldsymbol{v}_{b,k}-\boldsymbol{B}_{b}\boldsymbol{p}_{r}+\kappa\boldsymbol{\lambda}_{\boldsymbol{v}_{b,k}}\|^{2}+&\nonumber\\
&\|\tilde{\boldsymbol{v}}_{b,k}-\boldsymbol{v}_{b,k}^{H}\tilde{\boldsymbol{\Sigma}}\boldsymbol{B}_{k}+\kappa\boldsymbol{\lambda}_{\tilde{\boldsymbol{v}}_{b,k}}\|^{2}.&\label{pro50}
\end{align}
Based on (\ref{pro50}) $\boldsymbol{\mu}_{k}$ and $\boldsymbol{v}_{b,k}$ are given by
\begin{align}
&\boldsymbol{\mu}_{k}=(\boldsymbol{B}_{t,k^{\prime}}\tilde{\boldsymbol{w}}_{k^{\prime}}-\kappa\boldsymbol{\lambda}_{\boldsymbol{\mu}_{k^{\prime}}}+(\tilde{\boldsymbol{\mu}}_{k^{\prime}}+\kappa\boldsymbol{\lambda}_{\tilde{\boldsymbol{\mu}}_{k^{\prime}}})^{H}\boldsymbol{B}_{r}^{H}\boldsymbol{\Sigma}_{k})\nonumber\\
&(\boldsymbol{I}+(\boldsymbol{B}_{r}^{H}\boldsymbol{\Sigma}_{k})^{H}\boldsymbol{B}_{r}^{H}\boldsymbol{\Sigma}_{k})^{-*},\nonumber\\
&\boldsymbol{v}_{b,k}=(\boldsymbol{B}_{b}\boldsymbol{p}_{r}+\kappa\boldsymbol{\lambda}_{\boldsymbol{v}_{b,k}}+(\tilde{\boldsymbol{v}}_{b,k}+\kappa\boldsymbol{\lambda}_{\tilde{\boldsymbol{v}}_{b,k}})^{H}(\tilde{\boldsymbol{\Sigma}}\boldsymbol{B}_{k})^{H})\nonumber\\
&(\boldsymbol{I}+\tilde{\boldsymbol{\Sigma}}\boldsymbol{B}_{k}(\tilde{\boldsymbol{\Sigma}}\boldsymbol{B}_{k})^{H})^{-*}.\label{pro51}
\end{align}
For the \textbf{sub-block~7}, the subproblem with respect to $\boldsymbol{p}_{k}$ is given by
\begin{align}
\min_{\boldsymbol{p}_{k}}&\|\angle\boldsymbol{A}(\boldsymbol{p}_{k})-\angle\boldsymbol{B}_{k}+\kappa\boldsymbol{\lambda}_{\boldsymbol{B}_{k}}\|^{2}.&\label{pro52}
\end{align}
Since $\boldsymbol{B}_{k}(l,n_{1})$ satisfies constant modulus-constrained, problem (\ref{pro52}) is rewritten as
\begin{align}
&\mathcal{J}(\boldsymbol{p}_{k,n_{1}})=\sum\nolimits_{n_{1}=1}^{N_{u}}(\sum\nolimits_{l=1}^{L}|\frac{2\pi}{\lambda}(\boldsymbol{\pi}_{k}^{l})^{T}\boldsymbol{p}_{k,n_{1}}-\angle\boldsymbol{B}_{k}(l,n_{1})|^{2}\nonumber\\
&+\sum\nolimits_{n_{2}=1}^{N_{u}}\|\tilde{\boldsymbol{p}}_{k,n_{1}}-(\boldsymbol{p}_{k,n_{1}}-\boldsymbol{p}_{k,n_{2}})+\kappa\boldsymbol{\lambda}_{\tilde{\boldsymbol{p}}_{k,n_{1}}}\|^{2}).\label{pro53}
\end{align}
Based on the first-order optimal condition, $\boldsymbol{p}_{k,n_{1}}$ is given by
\begin{align}
&(\boldsymbol{I}+\sum_{l=1}^{L}\frac{4\pi^{2}}{\lambda^{2}}\boldsymbol{\pi}_{k}^{l}(\boldsymbol{\pi}_{k}^{l})^{T})\boldsymbol{p}_{k,n_{1}}-(\tilde{\boldsymbol{p}}_{k,n_{1}}+\sum_{n_{2}=1}^{N_{u}}\boldsymbol{p}_{k,n_{2}}+\nonumber\\
&\kappa\boldsymbol{\lambda}_{\tilde{\boldsymbol{p}}_{k,n_{1}}})-\sum_{l=1}^{L}\frac{2\pi}{\lambda}\boldsymbol{\pi}_{k}^{l}\angle\boldsymbol{B}_{k}(l,n_{1})=0.\label{pro54}
\end{align}
The solution of the equation (\ref{pro54}) is written as (\ref{pro56}) at the top of this page, in which $\mathcal{C}_{t}^{max}$ and $\mathcal{C}_{t}^{min}$ are the maximum mobile region and minimum mobile region of the MA position coordinate $\boldsymbol{p}_{k,n_{1}}$, respectively. Similarly, we can obtain the optimal solution for $\boldsymbol{p}_{r,m_{1}}$ using the same approach.
\begin{figure*}
\begin{align}
\boldsymbol{p}_{k,n_{1}}=\left\{\begin{matrix}
(\boldsymbol{I}+\sum_{l=1}^{L}\frac{4\pi^{2}}{\lambda^{2}}\boldsymbol{\pi}_{k}^{l}(\boldsymbol{\pi}_{k}^{l})^{T})^{-1}((\tilde{\boldsymbol{p}}_{k,n_{1}}+\sum_{n_{2}=1}^{N_{u}}\boldsymbol{p}_{k,n_{2}}+\kappa\boldsymbol{\lambda}_{\tilde{\boldsymbol{p}}_{k,n_{1}}})+\sum_{l=1}^{L}\frac{2\pi}{\lambda}\boldsymbol{\pi}_{k}^{l}\angle\boldsymbol{B}_{k}(l,n_{1})),&~\textrm{if}~ \boldsymbol{p}_{k,n_{1}}\in\mathcal{C}_{t}\\
\mathcal{C}_{t}^{max}, &~\textrm{else}~\textrm{if}~J(\mathcal{C}_{t}^{max})<J(\mathcal{C}_{t}^{min})\\
\mathcal{C}_{t}^{min}, &~\textrm{else}\\
\end{matrix}\right.,\label{pro56}
\end{align}
\hrulefill
\end{figure*}

\subsection{Variables in \textbf{Block~3} update}
In this subsection, we divide the variables in \textbf{Block~3} into $2$ sub-blocks. \{$\tilde{\boldsymbol{w}}_{k}$,$\tilde{\boldsymbol{\mu}}_{k^{\prime}}$,$\boldsymbol{u}_{J}$\} is called \textbf{sub-block~1}. \{$\Gamma_{k}$\} is called \textbf{sub-block~2}.

For the \textbf{sub-block~1}, the subproblem with respect to $\tilde{\boldsymbol{w}}_{k}$ is denoted as
\begin{align}
\min_{\tilde{\boldsymbol{w}}_{k}}&~\|\boldsymbol{\mu}_{k}-\boldsymbol{B}_{k}\tilde{\boldsymbol{w}}_{k}+\kappa\boldsymbol{\lambda}_{\boldsymbol{\mu}_{k}}\|^{2}+\|\boldsymbol{w}_{k}-\tilde{\boldsymbol{w}}_{k}+\kappa\boldsymbol{\lambda}_{\boldsymbol{w}_{k}}\|^{2}.&\label{pro57}
\end{align}
According to the first-order optimal condition, $\tilde{\boldsymbol{w}}_{k}$ is obtained
\begin{align}
\tilde{\boldsymbol{w}}_{k}=(\boldsymbol{B}_{k}^{H}(\boldsymbol{\mu}_{k}+\kappa\boldsymbol{\lambda}_{\boldsymbol{\mu}_{k}})+(\boldsymbol{w}_{k}+\kappa\boldsymbol{\lambda}_{\boldsymbol{w}_{k}}))(\boldsymbol{I}+\boldsymbol{B}_{k}^{H}\boldsymbol{B}_{k})^{-*}.\label{pro58}
\end{align}
Similarly, $\tilde{\boldsymbol{\mu}}_{k^{\prime}}$ and $\boldsymbol{u}_{J}$ are denoted as
\begin{align}
&\tilde{\boldsymbol{\mu}}_{k^{\prime}}=(\tilde{\boldsymbol{u}}_{t}^{H}(\tilde{u}_{k,k^{\prime}}+\kappa\lambda_{\tilde{u}_{k,k^{\prime}}})+(\kappa\boldsymbol{\lambda}_{\tilde{\boldsymbol{\mu}}_{k^{\prime}}}-\boldsymbol{B}_{r}^{H}\boldsymbol{\Sigma}_{k}\boldsymbol{\mu}_{k^{\prime}}))\nonumber\\
&(\boldsymbol{I}+\tilde{\boldsymbol{\mu}}_{k}^{H}\tilde{\boldsymbol{\mu}}_{k})^{-1},\nonumber\\
&\boldsymbol{u}_{J}=(\boldsymbol{f}^{H}(\boldsymbol{u}_{k}+\kappa\lambda_{\boldsymbol{u}_{k}})+(\kappa\boldsymbol{\lambda}_{\tilde{\boldsymbol{\mu}}_{k^{\prime}}}-\boldsymbol{B}_{r}^{H}\boldsymbol{\Sigma}_{k}\boldsymbol{\mu}_{k^{\prime}}))\nonumber\\
&(\boldsymbol{I}+\boldsymbol{f}_{k}^{H}\boldsymbol{f}_{k})^{-1}.\label{pro59}
\end{align}
For the \textbf{sub-block~1}, the subproblem with respect to $\Gamma_{k}$ is given by
\begin{subequations}
\begin{align}
\min_{\Gamma_{k}}&~|\Gamma_{k}-\tilde{\Gamma}_{k}+\kappa\lambda_{\Gamma_{k}}|^{2},&\label{pro60a}\\
\mbox{s.t.}~
&\tilde{\delta}_{k}^{2}+\frac{\Delta_{k}^{2}}{B^{2}\Gamma_{k}^{2}}\leq 2\alpha_{2,k}.&\label{pro60b}
\end{align}\label{pro60}%
\end{subequations}%
Similarly, we use Lagrange multiplier method to solve the problem in (\ref{pro58}), and the solution of $\Gamma_{k}$ and Lagrange multiplier $\kappa_{11,k}$ are expressed as
\begin{align}
&\Gamma_{k}(\kappa_{11,k})=\frac{\tilde{\Gamma}_{k}-\kappa\lambda_{\Gamma_{k}}}{2(1-\kappa_{11,k}(2\alpha_{2,k}-\tilde{\delta}_{k}^{2}))},\nonumber\\
&\kappa_{11,k}^{*}=\max\left\{0,\frac{2T\pm(\tilde{\Gamma}_{k}-\kappa\lambda_{\Gamma_{k}})\sqrt{2\alpha_{2,k}-\tilde{\delta}_{k}^{2}}}{(2\alpha_{2,k}-\tilde{\delta}_{k}^{2})}\right\}.\label{pro61}
\end{align}
The proposed algorithm is summarized in Algorithm~\ref{algo1}.
\begin{algorithm}%
\caption{Proposed Algorithm for Problem (\ref{pro14})} \label{algo1}
\hspace*{0.02in}{\bf Initialize:}
$\mathcal{X}^{(0)}$, $\rho^{(0)}>0$, $\boldsymbol{\lambda}^{(0)}$, $1<c<1$, $t=1$.\\
\hspace*{0.02in}{\bf Repeat:}~$t=t+1$.\\
$\{\mathbf{w}_{k},\tilde{\boldsymbol{f}}_{k}\}$ is computed based on (\ref{pro22}).\\ 
$\{\mathbf{B}_{k},\mathbf{B}_{r}, \mathbf{B}_{J}\}$ is computed based on (\ref{pro_25}).\\
\{$\gamma_{k}$,$\tilde{\gamma}_{k}$,$\tilde{\alpha}_{1,k}$,$\tilde{\alpha}_{2,k}$,$\bar{\delta}_{k}$,$\hat{\delta}_{k}$,$\tilde{\delta}_{k}$,$\tilde{\Psi}_{1,k}$,$\tilde{\Gamma}_{k}$,$\tilde{\nu}_{k}$\} is computed based on (\ref{pro27}), (\ref{pro28}).\\
\{$\tilde{\boldsymbol{p}}_{k,n_{1}}$,$\tilde{\boldsymbol{p}}_{r,m_{1}}$\} is computed based on (\ref{pro_29}), (\ref{pro31}).\\
\{$\tilde{u}_{k,k^{\prime}}$,$\tilde{\nu}_{k}$,$\boldsymbol{u}_{k}$\} is computed based on (\ref{pro33}).\\
$\gamma$ is computed based on (\ref{pro41}).\\
$\{\delta_{k}\}$ is computed based on 
(\ref{pro42}).\\
\{$\alpha_{1,k}$,$\alpha_{2,k}$,$\nu_{k}$\} is computed based on  (\ref{pro44}),
(\ref{pro45}).\\ 
\{$\Psi_{1,k}$\} is computed based on (\ref{pro__46}).\\
\{$\boldsymbol{f}_{k}$,$\boldsymbol{\mu}_{k^{\prime}}$\} is computed based on (\ref{pro48}).\\
\{$\boldsymbol{\mu}_{k}$,$\boldsymbol{v}_{b,k}$\} is computed based on (\ref{pro51}).\\
\{$\tilde{\boldsymbol{w}}_{k}$,$\tilde{\boldsymbol{\mu}}_{k^{\prime}}$,$\boldsymbol{u}_{J}$\} is computed based on (\ref{pro58}), (\ref{pro59}).\\
$ \{\Gamma_{k}\}$ is computed based on (\ref{pro61}).\\
\hspace*{0.02in}{\bf if:}~$\|\boldsymbol{h}(\mathcal{X}^{t})\|_{\infty}\leq\epsilon_{2}^{(t)}$.\\
Updating $\boldsymbol{\lambda}^{(t+1)}$ based on (\ref{pro19}),
$\kappa^{(t+1)}=\epsilon^{(t)}$.\\
\hspace*{0.02in}{\bf else:}~
Updating $\boldsymbol{\lambda}^{(t+1)}=\boldsymbol{\lambda}^{(t)}$,
$\kappa^{(t+1)}=\kappa^{(t)}$.\\
\hspace*{0.02in}{\bf Until:} some termination criterion is met.\\
\end{algorithm}


\subsection{Complexity Analysis}
To analyze the computational complexity of the proposed algorithm, we primarily focus on the updates of 
\{$\boldsymbol{w}_{k}$, $\tilde{\boldsymbol{f}}_{k}$\}, \{$\boldsymbol{B}_{k}$, $\boldsymbol{B}_{r}$, $\boldsymbol{B}_{J}$\}, \{$\tilde{u}_{k,k^{\prime}}$,$\tilde{\nu}_{k}$,$\boldsymbol{u}_{k}$\}, \{$\tilde{\boldsymbol{p}}_{k,n_{1}}$,$\tilde{\boldsymbol{p}}_{r,m_{1}}$\}, \{$\boldsymbol{f}_{k}$,$\boldsymbol{\mu}_{k^{\prime}}$\}, \{$\boldsymbol{p}_{k}$,$\boldsymbol{p}_{r}$\} and \{$\tilde{\boldsymbol{w}}_{k}$,$\tilde{\boldsymbol{\mu}}_{k^{\prime}}$,$\boldsymbol{u}_{J}$\}, which are the main determinants of complexity. 
The complexity of updating 
\{$\boldsymbol{w}_{k}$, $\tilde{\boldsymbol{f}}_{k}$\} is $\mathcal{O}(N_{t}+N_{r})$. When updating the \{$\boldsymbol{B}_{k}$, $\boldsymbol{B}_{r}$, $\boldsymbol{B}_{J}$\}, the complexity of one iteration of the proposed BCD-type algorithm is determined by $\mathcal{O}(L^{2}N_{t}^{2}+L^{2}N_{r}^{2}+\tilde{L}^{2}N_{r}^{2})$. The complexity of updating 
\{$\tilde{u}_{k,k^{\prime}}$,$\tilde{\nu}_{k}$,$\boldsymbol{u}_{k}$\} depends on the bisection method used to search for the Lagrange multiplier. The number of iterations required is 
$\mathcal{O}(\log_{2}(\theta_{0,s}/\theta_{s}))$, where 
$\theta_{0,s}$ represents the initial interval size and 
$\theta_{s}$ denotes the tolerance. Therefore, we can conclude that the computational cost of solving this subproblem is approximately.
The complexity of updating 
\{$\tilde{\boldsymbol{p}}_{k,n_{1}}$,$\tilde{\boldsymbol{p}}_{r,m_{1}}$\} is $\mathcal{O}(2N_{t}+2N_{r})$. 
The complexity of updating 
\{$\boldsymbol{f}_{k}$,$\boldsymbol{\mu}_{k^{\prime}}$\} is $\mathcal{O}(N_{r}^{3}K+N_{t}^{3}K)$. The complexity of updating 
\{$\boldsymbol{p}_{k}$,$\boldsymbol{p}_{r}$\} is $\mathcal{O}(2N_{t}+2N_{r})$. The complexity of updating 
\{$\tilde{\boldsymbol{w}}_{k}$,$\tilde{\boldsymbol{\mu}}_{k^{\prime}}$,$\boldsymbol{u}_{J}$\} is $\mathcal{O}(N_{t}^{3}+N_{r}^{3}+N_{J}^{3})$. Consequently, the overall complexity of the proposed algorithm can be represented as $\mathcal{O}(T_{1}T_{2}(\log_{2}(\theta_{0,s}/\theta_{s})+5N_{t}+5N_{r}+L^{2}N_{t}^{2}+L^{2}N_{r}^{2}+\tilde{L}^{2}N_{r}^{2}+2N_{r}^{3}+2N_{t}^{3}+N_{J}^{3}))$, in which the maximum number of iterations for
the inner and outer loops are denoted by $T_{1}$ and $T_{2}$.

\section{Numerical Results}\label{V}

In the simulation, UEs are uniformly distributed around the BS, with distances randomly distributed within a circular ring with a radius of $60$~meters(m), while jammers are distributed within a circular ring with a radius of $30$~m. The BS is located at the center of the ring. We employ the channel model from equations (\ref{pro1}) and (\ref{pro4}), where each user has the same number of transmission and reception paths, i.e., $L_{k}=2$ and $\tilde{L}=8$. For each user, 
$\boldsymbol{\Sigma}_{k}=\mathrm\{\sigma_{k,1}, \sigma_{k,2},\cdots,\sigma_{k,L}\}$ and $\tilde{\boldsymbol{\Sigma}}=\mathrm\{\sigma_{1}, \sigma_{2},\cdots,\sigma_{\tilde{L}}\}$ are the diagonal matrix, with each diagonal element following a circularly symmetric complex Gaussian (CSCG) distribution
$\mathcal{CN}(0, c_{k}^{2}/L)$ and $\mathcal{CN}(0, c^{2}/\tilde{L})$, where $c_{k}^{2}=g_{0}d_{k}^{-\alpha}$ and $c^{2}=g_{0}d_{J}^{-\alpha}$
represent the expected channel power gain for user $k$ and jammer, $g_{0}$
denotes the expected value of the average channel power gain at a reference distance of $1$~m, and $\alpha$ is the path loss exponent. It should be noted that, for a fair comparison, the total power of the elements in the path-response matrix for UEs with different path numbers is the same, i.e., $\mathbb{E}\{\mathrm{tr}(\boldsymbol{\Sigma}_{k}^{H}\boldsymbol{\Sigma}_{k})\}=c^2_{k}$. Additionally, we assume that the total size of the computational tasks is set to $1\times 10^{7}$ bits for all UEs. The computational resources of the MEC server are $1\times 10^{8}$ bits/s, and the local capacity for all UEs is $0.4\times 10^{7}$ bits/s. The bandwidth $B$ is set as $50$~MHz, and the noise power level is set as $10^{-5}$~mW. The transmission power $P_{k}$ and the jammer power $P_{J}$ are set to $25$~dBm and $0$~dBm, respectively. For the proposed algorithm, the tolerance parameter is chosen as $\epsilon_{1}= 1 \times 10^{-3}$. The initial penalty parameter is set to $\kappa^{0}=2$, with $c= 0.6$. Additionally, we set $\epsilon_{2}^{0}=0.1$ and $\epsilon_{2}^{m+1}=0.7\epsilon_{2}^{m}$. To demonstrate the superior performance of the proposed algorithm, this paper compares it with several existing methods. Specifically, channel matching (CM)-based, time division multiple access (TDMA)-based, and local computing algorithms are given in\cite{10032173}, while methods alternating position selection (APS), receive MA (RMA), maximum channel
power (MCP) and FPA are given in\cite{10243545}.

\begin{figure}[htbp]
\centering
\begin{minipage}[t]{0.48\textwidth}
\centering
\includegraphics[width=1\textwidth, height=0.6\textwidth]{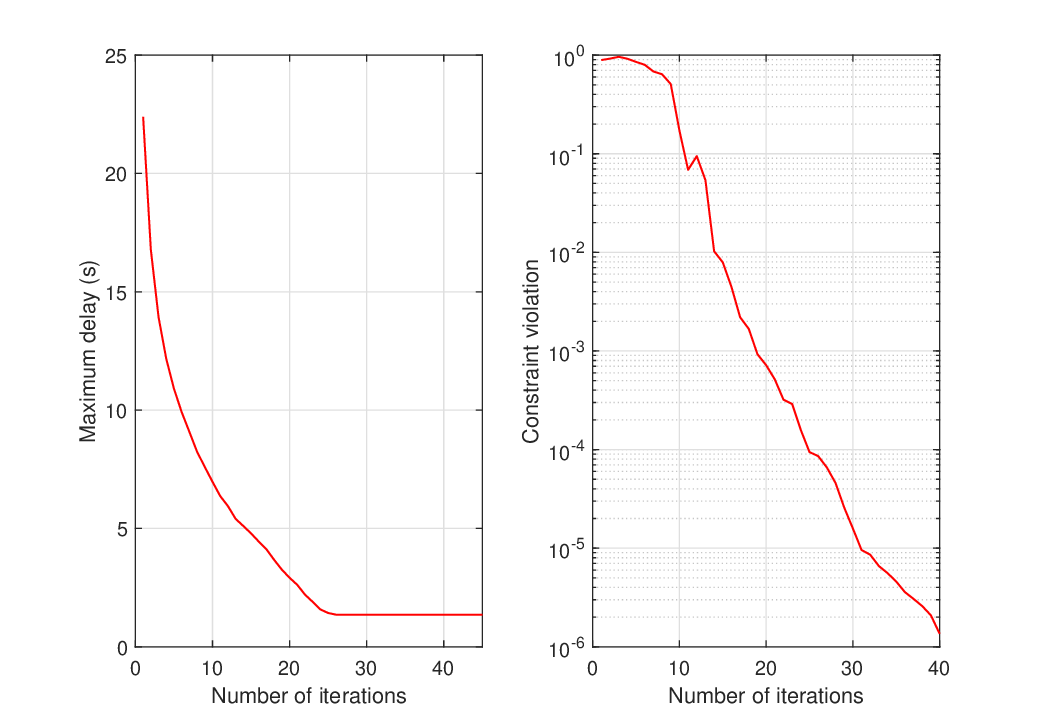}
\put(-180,-5){\small\textbf{(a)}}
\put(-70,-5){\small\textbf{(b)}}
\caption{(a) Maximum delay versus the number of outer iterations for the proposed algorithm. (b) Constraint violation versus the number of outer iterations for the proposed algorithm.}
\label{FIGURE2}
\end{minipage}
\begin{minipage}[t]{0.48\textwidth}
\centering
\includegraphics[width=1\textwidth, height=0.6\textwidth]{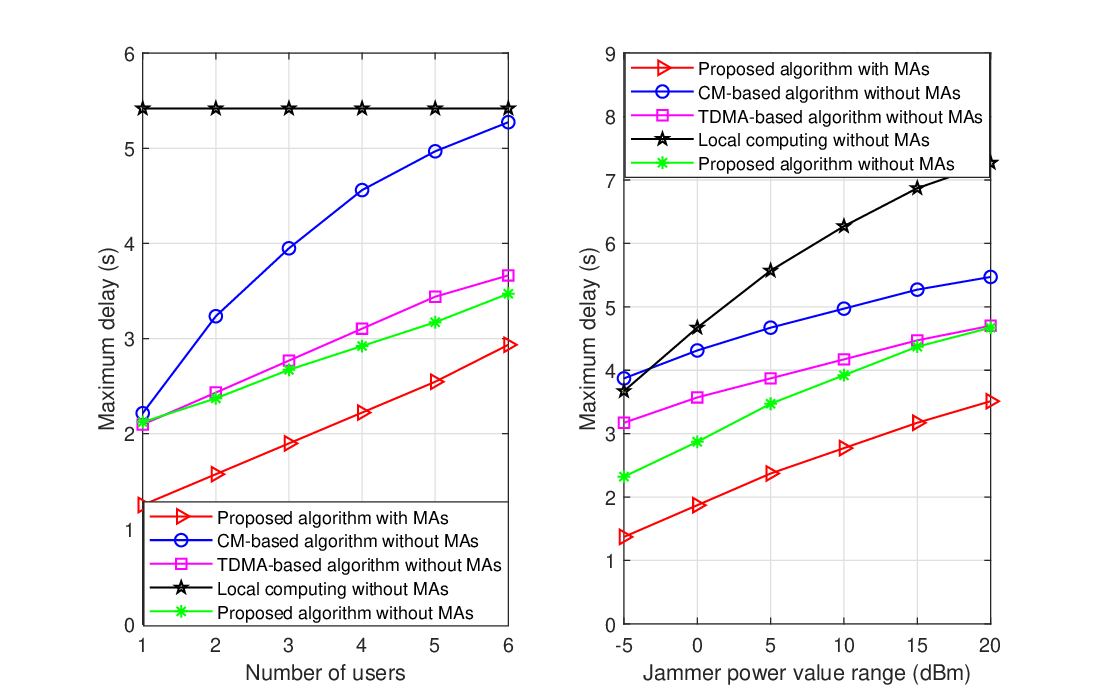}
\put(-180,-5){\small\textbf{(a)}}
\put(-70,-5){\small\textbf{(b)}}
\caption{(a) Maximum delay versus the number of UEs. (b) Maximum delay versus the jammer power. }
\label{FIGURE3}
\end{minipage}
\end{figure}

We first examine the convergence of the proposed algorithm, with simulation parameters set to $K=2$, $N_{t}=4$, $N_{r}=16$, $N_{J}=2$, $P_{k}=20$~dBm, and $P_{J}=5$~dBm. As shown in Fig.\ref{FIGURE2}(a), the proposed learning method converges within $24$ outer-loop iterations. Fig.\ref{FIGURE2}(b) illustrates the corresponding values of the constraint violation metric. It shows that the penalty term decreases to below $10^{-5}$ after $33$ iterations, indicating that the proposed algorithm based on SCA and PDD effectively addresses the equality constraints in the problem (\ref{pro14}).


Fig.\ref{FIGURE3}(a) compares the performance of four methods under varying numbers of UEs when $N_{t}=4$, $N_{r}=16$, $N_{J}=2$, $P_{k}=20$~dBm, and $P_{J}=5$~dBm. It can be seen that the performance of all methods improves as the number of UEs $K$ increases, with MA-based schemes outperforming nonMA methods. This is because MA provides a higher degree of freedom for performance optimization, leading to better performance gains when countering intelligent interference than traditional non-MA systems. Furthermore, as illustrated in Fig.\ref{FIGURE3}(a), in addition to local computation, the maximum latency of all algorithms increases with the number of UEs due to computational budget constraints, while the maximum latency for local computation remains constant. Furthermore, comparing the proposed algorithm with the CM-based and TDMA-based algorithms, we observe that as the number of UEs increases, the performance advantage of the proposed algorithm becomes more pronounced, demonstrating the benefits of the joint design. Lastly, the CM-based approach performs even worse for many UEs than the local computation method in this simulation scenario. This is due to the strong dependence of CM on channel conditions; When the distance increases, the channel conditions deteriorate, resulting in poorer CM performance, while the proposed algorithm provides the best performance among the methods analyzed even at greater distances. Fig.\ref{FIGURE3}(b) compares the performance of the four schemes under different jammer transmit power. The simulation parameters are set as $K=4$,
$N_{t}=4$, $N_{r}=16$, $N_{J}=2$, and $P_{k}=25$~dBm. Fig.\ref{FIGURE3}(b) indicates that as the jammer power increases, the system latency increases. This is because a higher jammer power requires more power resources to improve SINR, resulting in less power being allocated to MEC. We can also observe that the proposed algorithm and the baseline methods exhibit lower system latency across different power levels.

\begin{figure}[htbp]
\centering
\begin{minipage}[t]{0.48\textwidth}
\centering
\includegraphics[width=1\textwidth, height=0.6\textwidth]{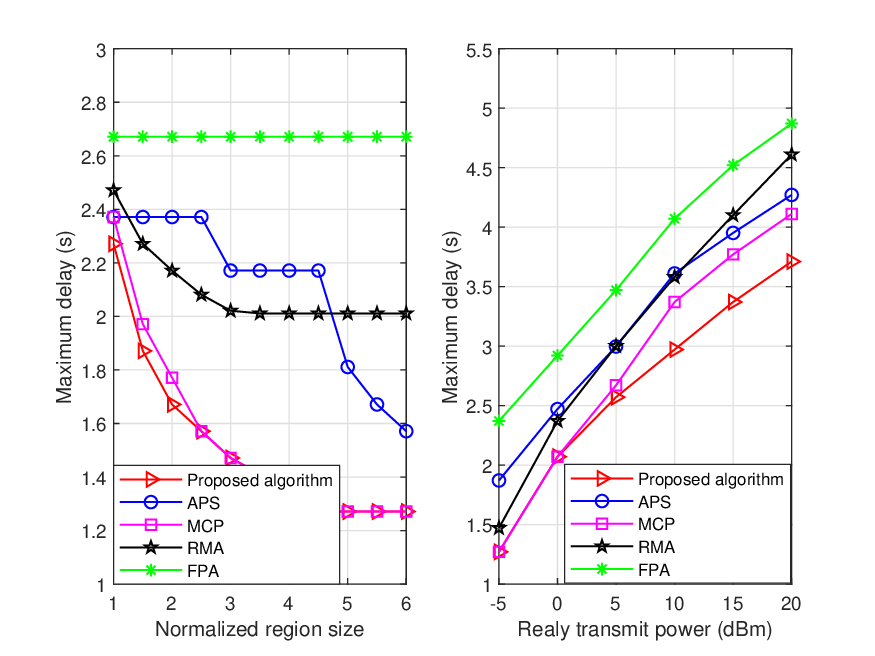}
\put(-180,-5){\small\textbf{(a)}}
\put(-70,-5){\small\textbf{(b)}}
\caption{(a) Maximum delay versus the MA mobile region.(a) Maximum delay versus the relay transmit power.}
\label{FIGURE4}
\end{minipage}
\begin{minipage}[t]{0.48\textwidth}
\centering
\includegraphics[width=1\textwidth, height=0.6\textwidth]{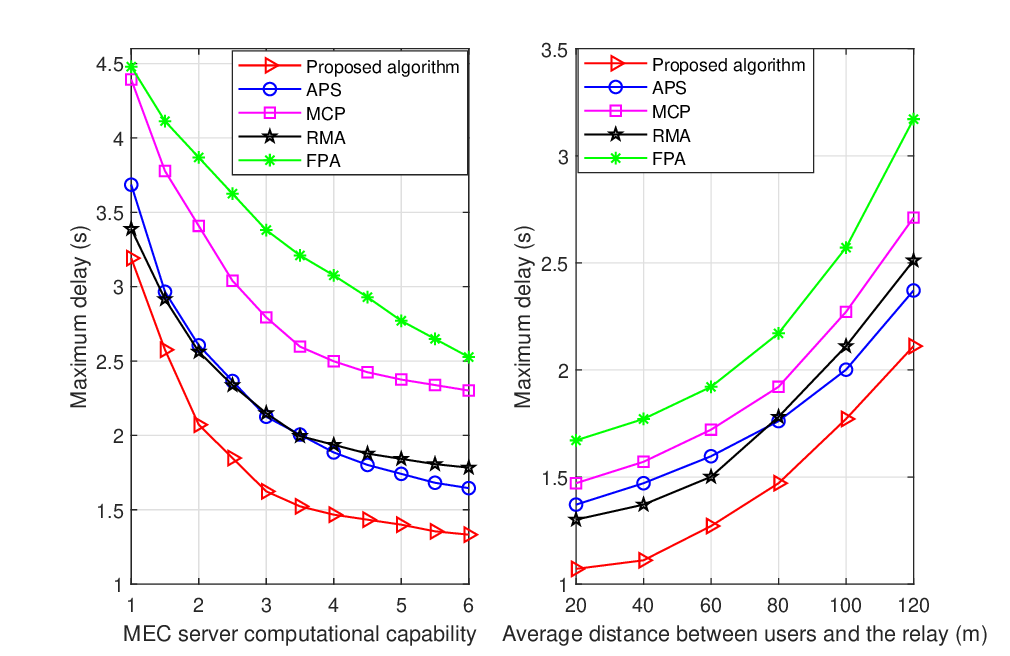}
\put(-180,-5){\small\textbf{(a)}}
\put(-70,-5){\small\textbf{(b)}}
\caption{(a) Maximum delay versus the MEC server computational capability.(a) Maximum delay versus the distance UEs and BS.}
\label{FIGURE5}
\end{minipage}
\end{figure}
In Fig.~\ref{FIGURE4}(a), simulation parameters set to $K=2$, $N_{t}=4$, $N_{r}=16$, $N_{J}=2$, $P_{k}=20$~dBm, and $P_{J}=5$~dBm. We present the relationship between the system maximum delay and the normalized area for the proposed MA-enabled MIMO system and benchmark schemes. It can be observed that the proposed algorithm and baseline schemes outperform the FPA system in terms of capacity, with performance gains increasing as the area size grows. Furthermore, when the normalized area exceeds $4$, all schemes converge, indicating that the maximum channel capacity of the MA-enabled MIMO system can be achieved with a limited transmission and reception area. In Fig.~\ref{FIGURE4}(b), simulation parameters set to $K=2$, $N_{t}=4$, $N_{r}=16$, $N_{J}=2$, and $P_{k}=20$~dBm. From Fig.~\ref{FIGURE4}(b), it is observed that as the jammer transmit power increases, the maximum system delay of all schemes increases. With higher jammer power, the system must allocate more resources to suppress interference to meet QoS requirements. However, as the jammer power increases, the performance loss of the MCP scheme is more significant. According to\cite{10243545}, the MCP scheme concentrates most of the channel power on the strongest eigenchannel, sacrificing the performance of other eigenchannels and resulting in poorer channel capacity than other schemes under high jammer power conditions.

In Fig.~\ref{FIGURE5}(a), simulation parameters set to $K=2$, $N_{t}=4$, $N_{r}=16$, $N_{J}=2$, $P_{k}=20$~dBm, and $P_{J}=5$~dBm.
Fig.~\ref{FIGURE5}(a) illustrates the relationship between the maximum system delay and the MEC computing capacity. Assuming there are three mobile UEs in the system, the computing capacity of the MEC server gradually increases. It can be observed that the system delay for both the PDD-based antenna design and other baseline algorithms decreases as the computing capacity of the MEC server increases, due to the availability of more resources on the MEC server. Moreover, when the MEC computing capacity becomes sufficiently high, the curve for the PDD-based algorithm approaches a fixed value, indicating that even with an increase in MEC computing resources, the maximum system delay does not decrease further. This is because the system delay is constrained by the mobility region of the MA and the availability of wireless resources. 
In Fig.~\ref{FIGURE5}(b), simulation parameters set to $K=2$, $N_{t}=4$, $N_{r}=16$, $N_{J}=2$, $P_{k}=20$~dBm, and $P_{J}=-5$~dBm. Fig.~\ref{FIGURE5}(b) shows that as the distance between UEs and BS increases, the maximum delay for both the PDD-based MA optimization algorithm and other baseline algorithms increases. This is due to the degradation of channel conditions with increasing distance, leading to longer transmission delays and poorer SINR for UEs. As the distance between UEs and the BS increases, the gap between the proposed joint design and other baseline algorithms becomes more pronounced, demonstrating that our proposed algorithm can achieve a more stable reduction in the maximum system delay.

\section{Conclusion}\label{VI}
This paper proposes using MA to enhance the anti-jamming performance of MEC systems. With the assistance of MA, the signal received by legitimate UEs is improved, while interference generated by jammers can be mitigated. Specifically, to address this challenge, a novel algorithm based on SCA and PDD is proposed, which jointly optimizes beamforming and MA positions under constraints of transmission power, computational resources, QoS for each user, and the MA's movable region. The proposed SCA and PDD-based algorithm operates through a dual-loop iterative process. We then discuss the convergence of the proposed algorithm and provide a detailed analysis of its computational complexity. Our simulation results demonstrate that the proposed algorithm outperforms traditional resource allocation schemes and MA algorithms in terms of system maximum delay.

\end{document}